\documentclass[11pt]{article}
\usepackage{amsmath,amsthm,latexsym,amssymb,amsfonts,epsfig}

\makeatletter
\@addtoreset{equation}{section}
\makeatother

\newtheorem{Theorem}{Theorem}[section]

\def\be{\begin{equation}}
\def\ee{\end{equation}}
\def\ba{\begin{eqnarray}}
\def\ea{\end{eqnarray}}

\newcommand\nn{\nonumber}
\newcommand\q{\quad}
\def\Nl{{\mathchoice
{\setbox0=\hbox{$\displaystyle\rm N$}\hbox{\hbox to0pt
{\kern0.4\wd0\vrule height0.9\ht0\hss}\box0}}
{\setbox0=\hbox{$\textstyle\rm N$}\hbox{\hbox to0pt
{\kern0.4\wd0\vrule height0.9\ht0\hss}\box0}}
{\setbox0=\hbox{$\scriptstyle\rm N$}\hbox{\hbox to0pt
{\kern0.4\wd0\vrule height0.9\ht0\hss}\box0}}
{\setbox0=\hbox{$\scriptscriptstyle\rm N$}\hbox{\hbox to0pt
{\kern0.4\wd0\vrule height0.9\ht0\hss}\box0}}}}
\def\Zl{{\mathchoice
{\setbox0=\hbox{$\displaystyle\rm Z$}\hbox{\hbox to0pt
{\kern0.4\wd0\vrule height0.9\ht0\hss}\box0}}
{\setbox0=\hbox{$\textstyle\rm Z$}\hbox{\hbox to0pt
{\kern0.4\wd0\vrule height0.9\ht0\hss}\box0}}
{\setbox0=\hbox{$\scriptstyle\rm Z$}\hbox{\hbox to0pt
{\kern0.4\wd0\vrule height0.9\ht0\hss}\box0}}
{\setbox0=\hbox{$\scriptscriptstyle\rm Z$}\hbox{\hbox to0pt
{\kern0.4\wd0\vrule height0.9\ht0\hss}\box0}}}}
\def\Ql{{\mathchoice
{\setbox0=\hbox{$\displaystyle\rm Q$}\hbox{\hbox to0pt
{\kern0.4\wd0\vrule height0.9\ht0\hss}\box0}}
{\setbox0=\hbox{$\textstyle\rm Q$}\hbox{\hbox to0pt
{\kern0.4\wd0\vrule height0.9\ht0\hss}\box0}}
{\setbox0=\hbox{$\scriptstyle\rm Q$}\hbox{\hbox to0pt
{\kern0.4\wd0\vrule height0.9\ht0\hss}\box0}}
{\setbox0=\hbox{$\scriptscriptstyle\rm Q$}\hbox{\hbox to0pt
{\kern0.4\wd0\vrule height0.9\ht0\hss}\box0}}}}
\def\Rl{{\mathchoice
{\setbox0=\hbox{$\displaystyle\rm R$}\hbox{\hbox to0pt
{\kern0.4\wd0\vrule height0.9\ht0\hss}\box0}}
{\setbox0=\hbox{$\textstyle\rm R$}\hbox{\hbox to0pt
{\kern0.4\wd0\vrule height0.9\ht0\hss}\box0}}
{\setbox0=\hbox{$\scriptstyle\rm R$}\hbox{\hbox to0pt
{\kern0.4\wd0\vrule height0.9\ht0\hss}\box0}}
{\setbox0=\hbox{$\scriptscriptstyle\rm R$}\hbox{\hbox to0pt
{\kern0.4\wd0\vrule height0.9\ht0\hss}\box0}}}}
\def\Cl{{\mathchoice
{\setbox0=\hbox{$\displaystyle\rm C$}\hbox{\hbox to0pt
{\kern0.4\wd0\vrule height0.9\ht0\hss}\box0}}
{\setbox0=\hbox{$\textstyle\rm C$}\hbox{\hbox to0pt
{\kern0.4\wd0\vrule height0.9\ht0\hss}\box0}}
{\setbox0=\hbox{$\scriptstyle\rm C$}\hbox{\hbox to0pt
{\kern0.4\wd0\vrule height0.9\ht0\hss}\box0}}
{\setbox0=\hbox{$\scriptscriptstyle\rm C$}\hbox{\hbox to0pt
{\kern0.4\wd0\vrule height0.9\ht0\hss}\box0}}}}
\def\Hl{{\mathchoice
{\setbox0=\hbox{$\displaystyle\rm H$}\hbox{\hbox to0pt
{\kern0.4\wd0\vrule height0.9\ht0\hss}\box0}}
{\setbox0=\hbox{$\textstyle\rm H$}\hbox{\hbox to0pt
{\kern0.4\wd0\vrule height0.9\ht0\hss}\box0}}
{\setbox0=\hbox{$\scriptstyle\rm H$}\hbox{\hbox to0pt
{\kern0.4\wd0\vrule height0.9\ht0\hss}\box0}}
{\setbox0=\hbox{$\scriptscriptstyle\rm H$}\hbox{\hbox to0pt
{\kern0.4\wd0\vrule height0.9\ht0\hss}\box0}}}}
\def\Ol{{\mathchoice
{\setbox0=\hbox{$\displaystyle\rm O$}\hbox{\hbox to0pt
{\kern0.4\wd0\vrule height0.9\ht0\hss}\box0}}
{\setbox0=\hbox{$\textstyle\rm O$}\hbox{\hbox to0pt
{\kern0.4\wd0\vrule height0.9\ht0\hss}\box0}}
{\setbox0=\hbox{$\scriptstyle\rm O$}\hbox{\hbox to0pt
{\kern0.4\wd0\vrule height0.9\ht0\hss}\box0}}
{\setbox0=\hbox{$\scriptscriptstyle\rm O$}\hbox{\hbox to0pt
{\kern0.4\wd0\vrule height0.9\ht0\hss}\box0}}}}
\newcommand{\ci}{\mathcal I}

  \newcommand{\Ff}{\mathfrak{F}}
  
\newcommand{\fh}{\mathfrak{h}}  
       
  \newcommand{\Fj}{\mathfrak{J}}

\newcommand{\eps}{\epsilon}

\DeclareMathOperator{\MC}{\boldsymbol{\mathsf{M}}}
\DeclareMathOperator{\MCO}{\boldsymbol{\widehat{\mathsf{M}}}}

\DeclareMathOperator{\MCW}{\rm Master\;\;Constraint}
\DeclareMathOperator{\MCOW}{\rm Master\;\;Constraint\;\;
Operator}
\DeclareMathOperator{\MCPW}{\rm Master\;\;Constraint\;\;Programme}

\title{Testing the\\ $\MCPW$\\ for Loop Quantum Gravity\\
IV. Free Field Theories}
\author{B.
Dittrich\thanks{dittrich@aei.mpg.de, bdittrich@perimeterinstitute.ca},
T. Thiemann\thanks{thiemann@aei.mpg.de, tthiemann@perimeterinstitute.ca}\\
\\
Albert Einstein Institut, MPI f. Gravitationsphysik\\
Am M\"uhlenberg 1, 14476 Potsdam, Germany\\
\\
and\\
\\
Perimeter Institute for Theoretical Physics \\
31 Caroline Street North, Waterloo, ON N2L 2Y5, Canada}
\date{{\small Preprint AEI-2004-119}}

\begin{document}

\maketitle

\begin{abstract}
This is the fourth paper in our series of five in which we test the Master
Constraint Programme for solving the Hamiltonian constraint in Loop
Quantum Gravity. We now move on to free field theories with constraints,
namely Maxwell theory and linearized gravity. Since the Master constraint
involves squares of constraint operator valued distributions, one has
to be very careful in doing that and we will see that the full flexibility
of the Master Constraint Programme must be exploited in order to arrive at
sensible results.
 \end{abstract}

\newpage

\tableofcontents

\section{Introduction}
\label{s1}

We continue our test of the Master Constraint Programme
\cite{7.0} for Loop Quantum Gravity (LQG) \cite{1.1,7.2,7.3}
which we started in the companion papers \cite{I,II,III} and will continue
in \cite{V}. The Master Constraint Programme is a new idea to improve on
the current situation with the Hamiltonian constraint operator for LQG
\cite{7.1}. In short, progress on the solution of the Hamiltonian
constraint has been slow because of a technical reason: the Hamiltonian
constraints
themselves are not spatially diffeomorphism invariant. This means that one
cannot first solve the spatial diffeomorphism constraints and then
the Hamiltonian constraints because the latter do not preserve the
space of solutions to the spatial diffeomorphism constraint
\cite{8.2}. On the other hand, the space of solutions
to the spatial diffeomorphism constraint \cite{8.2} is relatively easy to
construct starting from the spatially diffeomorphism invariant
representations on which LQG is based \cite{7.4} which are therefore
very natural to use and, moreover, essentially unique. Therefore one would
really like to keep these structures. The Master Constraint
Programme removes that technical obstacle by replacing the Hamiltonian
constraints by a single Master Constraint which is a spatially
diffeomorphism invariant integral of squares of the individual
Hamiltonian constraints which encodes all the necessary information about
the constraint surface and the associated invariants. See e.g.
\cite{7.0,I} for a full discussion of
these issues. Notice that the idea of squaring constraints is not new,
see e.g. \cite{2.1}, however, our concrete implementation is new and also
the Direct Integral Decomposition (DID) method for solving them, see
\cite{7.0,I} for all the details.

The Master Constraint for four
dimensional General Relativity will appear in \cite{8.1} but before we
test its semiclassical limit, e.g. using the methods of \cite{8.3,8.5}
and try to solve it by DID methods we want to test the programme in the
series of papers \cite{I,II,III,V}.
In the previous papers we focussed on finite dimensional
systems of various degrees of complexity. This time we face free quantum
fields squarely and will have to worry about the associated ultraviolet
singularities. We study free Maxwell Theory and Linearized Gravity
(formulated in connection variables) both of which are models
with an infinite number of Abelean first class constraints linear in the
momenta.

\section{Maxwell Theory}
\label{s6}

The canonical formulation of Maxwell Theory on $\Rl^4$ consists of an
infinite -- dimensional phase space $\cal M$ with canonically conjugate
coordinates $(A_a,E^a)$ and symplectic structure
\be \label{6.1}
\{A_a(x),A_b(y)\}=\{E^a(x),E^b(y)\}=0,\;\;
\{E^a(x),A_b(y)\}=e^2 \delta^a_b \delta(x,y)
\ee
where $e$ is the electric charge and we are using units so that
$\alpha=\hbar e^2$ is the Feinstrukturkonstante. In particular, the
$U(1)$ connection $A$ has units of cm$^{-1}$ while the electric field
$E$ has units of cm$^{-2}$.

The clean mathematical description of $\cal M$ models $\cal M$ on a Banach
space $\cal E$ \cite{6.1}. Here we will not need all the details of
this framework and it suffices to specify the fall -- off conditions of
$A,E$ at spatial infinity. Namely, in order that the canonical classical
action principle be well-defined, that is, the action be functionally
differentiable, the fields $A,E$ respectively must fall off at spatial
infinity at least as $r^{-1},r^{-2}$ respectively.

Consider now the infinite number of Maxwell -- Gauss Constraints
\be \label{6.2}
G(\Lambda)=\int_{\Rl^3}\; d^3x \Lambda (\partial_a E^a)=:
<\Lambda,\partial\cdot E>_{\mathfrak{h}}
\ee
where $\Lambda$ is a smooth test function of rapid decrease and
$\mathfrak{h}=L_2(\Rl,d^3x)$. Consider a positive definite operator
$C$ on $\mathfrak{h}$ independent of $\cal M$ and define the associated
$\MCW$ by
\be \label{6.3}
\MC:=\frac{1}{2} <\partial\cdot E,C\cdot\partial\cdot E>_{\mathfrak{h}}
\ee
Obviously $\MC=0$ if and only if $\partial\cdot E=0$ a.e., that is, if and
only if $G(\Lambda)=0$ for all test functions $\Lambda$ of rapid decrease.
For the same reason for a twice differentiable function $O$ on $\cal M$
we have $\{\{\MC,O\},O\}_{\MC=0}=0$ if and only if $\{\partial\cdot
E,O\}_{\MC=0}=0$ a.e, hence if and only if $\{G(\Lambda),O\}_{\MC=0}$ for 
all test
functions of rapid decrease.

Recall that the Maxwell -- Hamiltonian is given by ($\propto$ means
equal to on the constraint surface)
\be \label{6.4}
H=\frac{1}{2 e^2} \int \;d^3x\; \delta_{ab}\; (E^a E^b+B^a B^b)
\propto \hbar \int\; d^3x\; \delta_{ab}\; \overline{z^a}\; P^\perp_{ab}\; 
z^b
\ee
where the tranversal projector is given by
\be \label{6.5}
(P_\perp \cdot f)_a=f_a-\partial_a \Delta^{-1} \partial^b f_b
\ee
and
\be \label{6.6}
z^a=\frac{1}{\sqrt{2\alpha}}
[\sqrt{-\Delta}^{1/2} A_a-i\sqrt{-\Delta}^{-1/2} E^a]
\ee
Notice that we have used the Minkowski background metric in various
places which distinguishes this model from the subsequent ones in 
\cite{V}.
In the subsequent models we will treat background independent theories
which excludes background dependent Fock space quantization methods
which we will employ for this model.

In order to solve the $\MCW$ for Maxwell Theory on the usual Fock space it
is
mandatory to choose a nontrivial
operator $C$ in order that $\MCO$ becomes a densely defined operator
defined via annihilation and creation operators.

Let $b_I$ be any orthonormal basis of $\mathfrak{h}$ consisting
of real valued, smooth functions of rapid decrease. The index set $\cal I$
in which the $I$ takes
values is countable and we could for instance choose the $b_I$ to be
Hermite functions. Next, consider the Hilbert space
$\mathfrak{h}_3:=\mathfrak{h}^3$ with scalar product
$<f,f'>_{\mathfrak{h}_3}:=\int\; d^3x\;\delta^{ab}\; \overline{f_a} f'_b$.
Let us define the functions
\ba \label{6.7}
b'_I &:=& \sqrt{-\Delta}^{3/2} b_I
\nonumber\\
b^{(3)\prime}_{Ia}&:=& \sqrt{-\Delta}^{1/2} \partial_a b_I
\nonumber\\
b^{(3)}_{Ia}&:=& \sqrt{-\Delta}^{-1} \partial_a b_I
\ea
Notice the relations
\ba \label{6.8}
<b^{(3)\prime}_I,b^{(3)\prime}_J>_{\mathfrak{h}_3}
&=& <b'_I,b'_J>_{\mathfrak{h}}
\nonumber\\
<b^{(3)}_I,b^{(3)}_J>_{\mathfrak{h}_3}
&=& <b_I,b_J>_{\mathfrak{h}}=\delta_{IJ}
\ea
Let us complete the $b^{(3)}_I$ to an orthonormal basis
of $\mathfrak{h}_3$ by choosing some smooth, transversal and
orthonormal system $b^{(1)}_I,b^{(2)}_I$ of rapid decrease, that is,
$\partial^a b^{(j)}_{Ia}=0,\;j=1,2$. Now from (\ref{6.8}) it is clear that
the longitudinal vectors $b^{(3)\prime}_I$ do not form an orthonormal
system, but they can be, as elements of $\mathfrak{h}_3$, expanded in
terms
of the orthonormal basis $b^{(1)}_I,b^{(2)}_I,b^{(3)}_I$ where
only the expnasion coefficients for the $b^{(3)}_I$ are non-vanishing.

We thus find, using the completeness relation several times
\ba \label{6.9}
\MC&=&
\frac{1}{2} <\partial\cdot E,C\cdot\partial\cdot E>_{\mathfrak{h}}
\nonumber\\
&=&\frac{1}{2}
\sum_{I,J}
<\partial\cdot E,b_I>_{\mathfrak{h}}\;\;
<b_I,C\cdot b_J>_{\mathfrak{h}}\;\;
<b_J,\partial\cdot E>_{\mathfrak{h}}
\nonumber\\
&=&\frac{\alpha}{4}
\sum_{I,J}
<[z-\bar{z}],b^{(3)\prime}_I>_{\mathfrak{h}_3}\;\;
<b_I,C\cdot b_J>_{\mathfrak{h}}\;\;
<b^{(3)\prime}_J,[z-\bar{z}]>_{\mathfrak{h}_3}
\nonumber\\
&=&\frac{\alpha}{4}
\sum_{M,N}\; [\sum_{I,J}\;\;
<b^{(3)}_M,b^{(3)\prime}_I>_{\mathfrak{h}_3}\;\;
<b_I,C\cdot b_J>_{\mathfrak{h}}\;\;
<b^{(3)\prime}_J,b^{(3)}_N>_{\mathfrak{h}_3}]
\times\nonumber\\
&&\times
<[z-\bar{z}],b^{(3)}_M>_{\mathfrak{h}_3} \;
<b^{(3)}_N,[z-\bar{z}]>_{\mathfrak{h}_3}
\nonumber\\
&=&\frac{\alpha}{4}
\sum_{M,N}\; [\sum_{I,J}\;\;
<\sqrt{-\Delta}^{3/2}b_M,b_I>_{\mathfrak{h}}\;\;
<b_I,C\cdot b_J>_{\mathfrak{h}}\;\;
<b_J,\sqrt{-\Delta}^{3/2} b_N>_{\mathfrak{h}}]
\times\nonumber\\
&&\times
<[z-\bar{z}],b^{(3)}_M>_{\mathfrak{h}_3} \;
<b^{(3)}_N,[z-\bar{z}]>_{\mathfrak{h}_3}
\nonumber\\
&=&\frac{\alpha}{4}
\sum_{J,K}\;
<b_J,\sqrt{-\Delta}^{3/2} C\sqrt{-\Delta}^{3/2} b_K>_{\mathfrak{h}}\;\;
<[z-\bar{z}],b^{(3)}_J>_{\mathfrak{h}_3} \;
<b^{(3)}_K,[z-\bar{z}]>_{\mathfrak{h}_3}
\ea
If we would choose $C=\sqrt{-\Delta}^{-3}$ then (\ref{6.9}) would simplify
very much, however, this is impossible due to the boundary conditions
on $E$ which would imply that for this choice of $C$ the integral
(\ref{6.3}) diverges logarithmically. However, notice that the operator
$Q$ on $\mathfrak{h}$ defined by
\be \label{6.10}
Q=\sqrt{-\Delta}^{3/2}C\sqrt{-\Delta}^{3/2}
\ee
is positive definite.

We now come to the quantization of the system. We consider
the kinematical operator algebra $\mathfrak{A}$ generated from
annihilation operators
\be \label{6.12}
\hat{z}^{(j)}_J:=\widehat{<b^{(j)}_J,z>_{\mathfrak{h}_3}}
\ee
and the corresponding creation operators given by their adjoints which
are subject to the usual commutation relations
\be \label{6.12a}
[\hat{z}^j_J,\hat{z}^k_K]=[(\hat{z}^j_J)^\dagger,(\hat{z}^k_K)^\dagger]
=0,\;\;
[\hat{z}^j_J,(\hat{z}^k_K)^\dagger]=\alpha \delta^{jk}\delta_{JK}
\ee
We represent this algebra on the usual
kinematical Hilbert space ${\cal H}_{Kin}$ of Maxwell theory given by
the Fock space $\cal F$ generated from the cyclic vacuum
vector $\Omega$ defined by
\be \label{6.13}
\hat{z}^j_J\Omega=0
\ee
In terms of creation and annihilation operators, the $\MCOW$ becomes
\be \label{6.14a}
\MCO=\frac{\alpha}{4}
\sum_{J,K}\; Q^{JK}\;
[\hat{z}^3_J-(\hat{z}^3_J)^\dagger]^\dagger\;
[\hat{z}^3_K-(\hat{z}^3_K)^\dagger]
\ee
Notice that (\ref{6.13}) is not normal ordered, hence it is a non-trivial
question whether (\ref{6.13}) is even densely defined.
\begin{Theorem} \label{th6.1} ~~~\\
The $\MCOW$ (\ref{6.14a}) is densely defined if and only if $Q$ is a
trace class operator.
\end{Theorem}
Proof of theorem \ref{th6.1}:\\
Since the Fock space is the closure of the finite linear span of
finite excitations of the vacuum $\Omega$, $\MCO$ is densely defined
if and only if $\MCO\Omega$ has finite norm. We compute
\ba \label{6.14}
||\MCO\Omega||^2
&=&
(\frac{\alpha}{4})^2\sum_{J,K,M,N} Q_{JK} Q_{MN}
<[\hat{z}^3_J-(\hat{z}^3_J)^\dagger](\hat{z}^3_K)^\dagger \Omega,
[\hat{z}^3_M-(\hat{z}^3_M)^\dagger](\hat{z}^3_N)^\dagger \Omega>
\nonumber\\
&=&
(\frac{\alpha}{4})^2\sum_{J,K,M,N} Q_{JK} Q_{MN}
<\alpha\delta_{JK}-(\hat{z}^3_J)^\dagger(\hat{z}^3_K)^\dagger] \Omega,
\alpha\delta_{MN}-(\hat{z}^3_M)^\dagger(\hat{z}^3_N)^\dagger] \Omega>
\nonumber\\
&=&
(\frac{\alpha}{4})^2\sum_{J,K,M,N} Q_{JK} Q_{MN}
[\alpha^2\delta_{JK}\delta_{MN}+
<\Omega,\hat{z}^3_K \hat{z}^3_J (\hat{z}^3_M)^\dagger(\hat{z}^3_N)^\dagger
\Omega>
\nonumber\\
&=&
(\frac{\alpha}{2})^4\sum_{J,K,M,N} Q_{JK} Q_{MN}
[\delta_{JK}\delta_{MN}+\delta_{JM}\delta_{KN}+
\delta_{JN}\delta_{KM}]
\nonumber\\
&=&
(\frac{\alpha}{2})^4[2\mbox{Tr}(Q^2)+(\mbox{Tr}(Q))^2]
\ea
Both terms in the last line of (\ref{6.14}) must be finite since $Q$ is
symmetric (even positive). The first term in the last line of (\ref{6.14})
is finite if $Q$
is a Hilbert -- Schmidt (or nuclear) operator, the second if $Q$ is
trace class. Since every trace class  operator is nuclear, it is
necessary and sufficient that $Q$ be trace class.\\
$\Box$\\
\\
Now the algebra of trace class
operators comprises an ideal within the compact operators. Compact
operators are bounded, have pure point spectrum and every non-zero
eigenvalue has finite multiplicity, hence zero is an accumulation point
in their spectrum unless $Q$ is a finite rank operator. A possible choice
for $C$ would therefore be for example
\be \label{6.15}
C=\sqrt{-\Delta}^{-3/2}\; e^{\frac{l^2\Delta-||x||^2/l^2}{2}}
\sqrt{-\Delta}^{-3/2}
\ee
where $l$ is an arbitrary finite length scale. The exponential in
(\ref{6.15}) is nothing else than minus a three dimensional harmonic
oscillator operator, so its eigenvalues are
$\lambda_n:=\exp(-[\frac{3}{2}+n_1+n_2+n_3]),\;n_j\in\Nl_0$, hence its
trace is
explicitly $e^{-3/2} (1-e^{-1})^{-3}$ and that of its square
$e^{-3} (1-e^{-2})^{-3}$.
\begin{Theorem} \label{th6.2} ~~~\\
The choice (\ref{6.15}) makes the integral (\ref{6.3}) converge.
\end{Theorem}
Proof of theorem \ref{th6.2}:\\
We expand (\ref{6.3}) in terms of coherent states $\psi_z$ for the
Hamiltonian $H$ defined by
\be \label{6.16}
\psi_z=e^{-||z||^2/2} \;\;e^{z^a\hat{z}_a^\dagger}|0>
\ee
where
\be \label{6.17}
z^a=\frac{1}{\sqrt{2}}[x^a/l-il p_a],\;\;
\hat{z}_a=\frac{1}{\sqrt{2}}[x^a/l+l \partial/\partial x^a]
\ee
Using the overcompleteness relation for coherent states
\be \label{6.18}
\int_{\Cl^3} \frac{d^3z d^3\bar{z}}{\pi^3}
\psi_z <\psi_z,.>_{\mathfrak{h}}=\mbox{id}_{\mathfrak{h}}
\ee
we find
\ba \label{6.19}
\MC
&=& \frac{1}{2}
\int_{\Cl^3} \frac{d^3z d^3\bar{z}}{\pi^3}
\int_{\Cl^3} \frac{d^3z' d^3\bar{z'}}{\pi^3}
\;\times\nonumber\\ &&\times
<\sqrt{-\Delta}^{-3/2}\partial\cdot E,\psi_z>_{\mathfrak{h}}\;
<\psi_z, e^{-H}\psi_{z'}>_{\mathfrak{h}}\;
<\psi_{z'},\sqrt{-\Delta}^{-3/2}\partial\cdot E>_{\mathfrak{h}}
\nonumber\\
&\le& \frac{1}{2}
\int_{\Cl^3} \frac{d^3z d^3\bar{z}}{\pi^3}
\int_{\Cl^3} \frac{d^3z' d^3\bar{z'}}{\pi^3}
\;\times\nonumber\\ &&\times
|<E,\sqrt{-\Delta}^{-3/2} d\cdot\psi_z>_{\mathfrak{h}_3}|\;\;
[|<\psi_z, e^{-H}\psi_{z'}>_{\mathfrak{h}}|]\;\;
|<\sqrt{-\Delta}^{-3/2} d\cdot\psi_{z'},E>_{\mathfrak{h}_3}|
\nonumber\\
&\le& \frac{||E||^2_{\mathfrak{h}_3}}{2}
\int_{\Cl^3} \frac{d^3z d^3\bar{z}}{\pi^3}
\int_{\Cl^3} \frac{d^3z' d^3\bar{z'}}{\pi^3}
\;\times\nonumber\\ &&\times
||\sqrt{-\Delta}^{-3/2} d\cdot\psi_z||_{\mathfrak{h}_3}\;\;
[|<\psi_z, e^{-H}\psi_{z'}>_{\mathfrak{h}}|]\;\;
||\sqrt{-\Delta}^{-3/2} d\cdot\psi_{z'}||_{\mathfrak{h}_3}
\nonumber\\
&=& \frac{||E||^2_{\mathfrak{h}_3}}{2}
\int_{\Cl^3} \frac{d^3z d^3\bar{z}}{\pi^3}
\int_{\Cl^3} \frac{d^3z' d^3\bar{z'}}{\pi^3}
\;\times\nonumber\\ &&\times
\sqrt{<\psi_z,\sqrt{-\Delta}^{-1}\psi_z>_{\mathfrak{h}}}\;\;
[|<\psi_z, e^{-H}\psi_{z'}>_{\mathfrak{h}}|]\;\;
\sqrt{<\psi_{z'},\sqrt{-\Delta}^{-1}\psi_{z'}>_{\mathfrak{h}}}\;
\ea
where in the third step we have used the Schwartz inequality.

Since the
classical electric field energy $\frac{||E||^2_{\mathfrak{h}_3}}{2}$
converges, it suffices to estimate the integrals in (\ref{6.19}).
Using the expansion of the coherent states into energy eigenfunctions
it is easy to see that
\ba \label{6.20}
&&
|<\psi_z, e^{-H}\psi_{z'}>_{\mathfrak{h}}|
=e^{-3/2}
|\exp(-\frac{1}{2}[z_a\bar{z}_a+z'_a\bar{z}'_a-2\bar{z}_a z_a'/e])|
\nonumber\\
&=& e^{-3/2}
\exp(-\frac{1}{2e}\sum_a |z_a-z'_a|^2)\;\;
\exp(-\frac{1}{2}(1-\frac{1}{e}) [z_a\bar{z}_a+z'_a\bar{z}'_a])
\nonumber\\
&\le& e^{-3/2}
\exp(-\frac{1}{2}(1-\frac{1}{e}) [z_a\bar{z}_a+z'_a\bar{z}'_a])
\ea
Next, using the Fourier transform we get (up to a constant phase)
\be \label{6.21}
\tilde{\psi}_z(k)=\int d^3x e^{-ikx} \psi_z(x)=
(2l\sqrt{\pi})^{3/2}\exp(-||p||^2 l^2/2+||k||^2 l^2/2-il k\cdot z)
\ee
and have by using polar coordinates
\ba \label{6.22}
&&<\psi_z,\sqrt{-\Delta}^{-1}\psi_z>_{\mathfrak{h}}
=
(2l\sqrt{\pi})^3
e^{-||p||^2 l^2} \int \frac{d^3k}{||k||(2\pi)^3}
e^{-||k||^2 l^2-2l^2 k\cdot p}
\nonumber\\
&=&2\pi l
\sqrt{\pi}^{-3}
e^{-||p||^2 l^2} \int_0^\infty r dr e^{-r^2} \int_{-1}^1 dt
e^{-2l ||p|| r t}
\nonumber\\
&=&\pi
\sqrt{\pi}^{-3}\frac{1}{||p||}
\int_0^\infty  dr [e^{-(r- l||p||)^2}-e^{-(r+l||p||)^2}]
\nonumber\\
&=&\frac{1}{||p||}
\int_{-l ||p||}^{l ||p||} \frac{dr}{\sqrt{\pi}} e^{-r^2}
\nonumber\\
&\le & \frac{1}{||p||}
\ea
Thus we can finish our estimate by
\ba \label{6.23}
\MC
&\le& \frac{||E||^2_{\mathfrak{h}_3}}{2} \frac{e^{-3/2}}{\pi ^6}
[\int d^3x e^{-\frac{1}{2l^2}(1-\frac{1}{e})||x||^2}]^2\;\;
[\int \frac{d^3p}{\sqrt{||p||}}
e^{-\frac{l^2}{2}(1-\frac{1}{e})||p||^2}]^2
\nonumber\\
&=& \frac{||E||^2_{\mathfrak{h}_3}}{2} \frac{l e^{-3/2}}{\pi ^6}
\sqrt{1-\frac{1}{e}}^{-11}
[\int d^3x e^{-\frac{1}{2}||x||^2}]^2\;\;
[\int \frac{d^3k}{\sqrt{||k||}}
e^{-\frac{1}{2}||p||^2}]^2
\nonumber\\
&=& l\frac{||E||^2_{\mathfrak{h}_3}}{2} \frac{ e^{-3/2}}{\pi ^6}
\sqrt{1-\frac{1}{e}}^{-11} (2\pi)^3 (4\pi)^2
[\int_0^\infty r^{3/2} dr e^{-\frac{1}{2}r^2}]^2
\nonumber\\
&=& l\frac{||E||^2_{\mathfrak{h}_3}}{2} \frac{e^{-3/2}\sqrt{2} 2^7}{\pi}
\sqrt{1-\frac{1}{e}}^{-11} \Gamma(5/4)^2
\ea
where the last integral resulted in a $\Gamma-$function.\\
$\Box$\\
\\
In what follows we will not need the specifics of $Q$, we choose any trace
class operator such that (\ref{6.9}) converges.

Having made sure that both the classcial and quantum $\MCW$ are
well-defined
we can proceed to the solution of both the classical and the quantum
problem.


\subsection{Physical Hilbert Space}
\label{s6.2}

The full Fock space $\cal F$ can be conveniently written in the form
${\cal F}={\cal F}^\perp\otimes {\cal F}^\parallel$ where
${\cal F}^\parallel={\cal F}^{(3)}$ contains the longitudinal excitations
while ${\cal F}^\perp={\cal F}^{(1)}\otimes {\cal F}^{(2)}$ constains the
transversal ones. The longitudinal
Hilbert space in turn acquires the direct sum structure
\be \label{6.24}
{\cal F}^\parallel=\overline{
\oplus_{\{n^{(3)}_I\}_{I\in {\cal I}};\;\sum_I n^{(3)}_I<\infty}
\;\;\;{\cal F}^\parallel_{\{n^{(3)}\}}}
\ee
where the overline denotes closure and
the Hilbert spaces  ${\cal F}^\parallel_{\{n^{(3)}\}}$ are one
dimensional and can be identified with the multiples of the vector
\be \label{6.25}
\prod_I \;[(\hat{z}^{(3)}_I)^\dagger]^{n^{(3)}_I}\;\;\Omega
\ee
For our purposes it is more convenient to write (\ref{6.24}) in a
different way: Denote by $S({\cal I})$ the set of all finite subsets
${\cal J}\subset {\cal I}$. For each ${\cal J}\in S({\cal I})$ consider
the subspace ${\cal F}^\parallel_{{\cal J}}$ of ${\cal F}^\parallel$
consisting of the closure of the finite linear span of vectors such that
$n^{(3)}_I\ge 0$ if $I\in{\cal J}$ and $n^{(3)}_I=0$ otherwise. The
Hilbert space ${\cal F}^\parallel$ is the inductive limit of the family
of Hilbert
spaces ${\cal F}^\parallel_{{\cal J}}$ where directed set $S({\cal I})$ is
equipped with the inclusion as partial order.

Following the general programme of the direct integral decomposition
(DID) we should now simply find a cyclic system $\Omega_n$ of
mutually orthogonal $C^\infty-$vectors for $\MCO$ so that the closures
of the finite linear spans of vectors of the form $\MCO^N\Omega_n$ are
mutually orthogonal for different $n$. While we know that this can be done
in principle, it turns out to be a hard problem and we were not able to
find an explicit system $\Omega_n$. We leave it to the ambitious reader
to come up with such an explicit solution for the present case.
In what follows we therefore follow an
indirect strategy which is also useful for other complicated field
theoretic problems where it may be hard, if not impossible, to find an
explicit solution $\{\Omega_n\}$. \\
\\
It will be convenient to choose the $b_I$ to be
the eigenstates of the trace class operator $Q$ with positive eigenvalues
$Q_I$. For the example we discussed above the $b_I$ would be Hermite
functions. Then for any ${\cal J}\in S({\cal I})$ we define
\be \label{6.30}
\MCO_{{\cal J}}:=-(\frac{\alpha}{2})^4\sum_{I\in{\cal J}} Q_I
[\hat{z}^{(3)}_I-(\hat{z}^{(3)}_I)^\dagger]^2
\ee
One can check that the family of operators (\ref{6.30}) is {\it not} an
inductive family
with respect to the inductive limit structure introduced, basically
because it does not annihilate the vacuum. However, the following holds.
\begin{Theorem} \label{th6.3} ~~~\\
The family of operators (\ref{6.30}) converges strongly to $\MCO$
on the dense set of Fock states.
\end{Theorem}
Proof of theorem \ref{th6.3}:\\
To see this let $\psi\in{\cal F}^\parallel$ be a Fock state, that is,
a state which is a linear combination of states of the form (\ref{6.25}).
Then we find a minimal
${\cal J}_\psi\subset {\cal I}$ such that
$\psi\in{\cal F}^\parallel_{{\cal J}_\psi}$. Now consider any ${\cal J}$
with ${\cal J}_\psi\subset {\cal J}$. Then
\ba \label{6.32}
||(\MCO-\MCO_{{\cal J}})\psi||^2_{{\cal F}^\parallel} &=&
<\psi,(\MCO-\MCO_{{\cal J}})^2\psi>_{{\cal F}^\parallel}
\nonumber\\
&=&<\Omega,(\MCO-\MCO_{{\cal J}})^2\Omega>_{{\cal F}^\parallel}
\nonumber\\
&=&(\frac{\alpha}{2})^4\{
2\sum_{I\in{\cal I}-{\cal J}} Q_I^2
+[\sum_{I\in{\cal I}-{\cal J}} Q_I]^2\}
\ea
where in the second step we used that the operator $\MCO-\MCO_{{\cal J}}$
commutes with all the annihilation operators which create $\psi$ from
the vacuum $\Omega$ together with $||\psi||=1$. In the last line we used
the previous computation (\ref{6.14}).

Now since $Q$ is trace class, given $\epsilon>0,\;\psi\in{\cal
F}^\parallel$ we find
${\cal J}_1$ such that ${\cal J}_\psi\subset {\cal J}_1$ and such that
(\ref{6.32}) is
smaller than $\epsilon$. Since the series in (\ref{6.32}) is monotonously
decreasing as ${\cal J}\to {\cal I}$, this holds also for all
${\cal J}_1\subset {\cal J}$ which establishes the proof. We note that
the label ${\cal J}_1$ depends on $\epsilon$ but not on $\psi$ itself but
only on ${\cal J}_\psi$, hence the limit is partly uniform.\\
$\Box$\\
\\
The idea to solve the constraint now rests on the following observation.
\begin{Theorem} \label{thsa.1} ~\\
Let $a_n$ be a sequence of self -- adjoint operators with common dense
domain $D$ such that i) $a_n \psi\to a\psi$ for all $\psi\in D$ where $a$
is another self -- adjoint operator also defined densely on $D$. Suppose
ii) that $x,y\not\in
\sigma_{pp}(a)$. Then $s-\lim_{n\to \infty} E_n((x,y))=E((x,y))$.
\end{Theorem}
Proof of theorem \ref{thsa.1}:\\
Recall that $a_n\to a$ in the strong resolvent sense provided that
$R_z(a_n)\to R_z(a)$ strongly for any (and therefore all, see
theorem VIII.19 of \cite{RS}) $z\in \Cl$ with $\Im(z)\not=0$. Here
$R_z(a)=(a-z)^{-1}$ is the resolvent of $a$. By i)
and \cite{RS}, theorem viii.25a), $a_n\to a$ in the strong resolvent
sense. By ii) and \cite{RS} theorem viii.24b) the claim follows.\\
$\Box$\\
The theorem applies in particular when $a_n,a$ have purely continuous
spectrum in which case the convergence holds for all measurable sets.\\
\\
We can apply this theorem to our case because both assumptions i)
and ii) are satisfied. Namely, by theorem \ref{th6.3}) all the operators
$\MCO_{{\cal J}}$ are densely defined on the dense set of finite
linear combinations of Fock sates and converge there to $\MCO$ strongly.
Furthermore, the operators
$\hat{p}_I:=i[\hat{z}^{(3)}_I-(\hat{z}^{(3)}_I)^\dagger]$ are mutually
commuting, self -- adjoint and have absolutely continuous spectrum. Hence
also the $\MCO_{{\cal J}},\MCO$ are mutually commuting and self adjoint
(by the spectral theorem) and have absolutely continuous sectrum.
It follows that
$s-\lim_{{\cal J}\to {\cal I}} E_{{\cal J}}(B)=E(B)$ for any measurable
set $B$ where $E_{{\cal J}},\;E$ denote the p.v.m. of $\MCO_{{\cal J}}$
and $\MCO$ respectively.

Let $\Omega_0$ be any vector such that
$\mu_{\Omega_0}$ is of maximal type (not to be confused with the Fock
vacuum).
For instance it may be the vector
which actually defines the spectral measure $\mu$ underlying a direct
integral representation of ${\cal F}^\parallel$ subordinate to $E$.
We then have
\ba \label{6.32a}
\frac{d\mu_{\Psi,\Psi'}(0)}{d\mu_{\Omega_0}(0)}
&=& \lim_{x\to 0+} \frac{\mu_{\Psi,\Psi'}(x)}{\mu_{\Omega_0}(x)}
\nonumber\\
&=& \lim_{x\to 0+}
\frac{<\Psi,E((-\infty,x])\Psi'>}{<\Omega_0,E((-\infty,x])\Omega_0>}
=\frac{<\psi(0),\psi'(0)>_{{\cal
H}^\oplus_0}}{<\omega_0(0),\omega_0(0)>_{{\cal H}^\oplus_0}}
\ea
where the last equality holds $\mu-$a.e. and $\psi,\psi',\omega_0$ are the
direct integral representations of $\Psi,\Psi',\Omega_0$ respectively.

Since $\cal I$ is countable it is in bijection with $\Nl$ and thus
we take $I=1,2,..$ for simplicity of notation. Consider any $n,m$ and let
$\Psi_m,\Psi'_m,\Omega_{0m}$ be vectors which are finite linear
combinations
of Fock states with excitations at most up to label $I=m$ and let
$E_n$ be the p.v.m. of $\MCO_n:=(\alpha/2)^4\sum_{I=1}^n \hat{p}_I^2$.
We calculate with $\mu^n_{\Omega_0}(B):=<\Omega_0,E_n(B)\Omega_0>$ etc.
\ba \label{6.32b}
&&
\frac{\mu_{\Psi,\Psi'}(x)}{\mu_{\Omega_0}(x)}-
\frac{\mu^n_{\Psi_m,\Psi'_m}(x)}{\mu^n_{\Omega_{0m}}(x)}
\nonumber\\
&=&
[\frac{\mu_{\Psi,\Psi'}(x)}{\mu_{\Omega_0}(x)}-
\frac{\mu^n_{\Psi,\Psi'}(x)}{\mu^n_{\Omega_0}(x)}]
+
[\frac{\mu^n_{\Psi,\Psi'}(x)}{\mu^n_{\Omega_0}(x)}]
-\frac{\mu^n_{\Psi_m,\Psi'_m}(x)}{\mu^n_{\Omega_{0m}}(x)}]
\nonumber\\
&=&
\frac{
\mu_{\Psi,\Psi'}(x)\mu^n_{\Omega_0}(x)-
\mu^n_{\Psi,\Psi'}(x)\mu_{\Omega_0}(x)
}
{\mu_{\Omega_0}(x)\mu^n_{\Omega_0}(x)}
+
\frac{
\mu^n_{\Psi,\Psi'}(x)\mu^n_{\Omega_{0m}}(x)-
\mu^n_{\Psi_m,\Psi'_m}(x)\mu^n_{\Omega_0}(x)
}
{\mu^n_{\Omega_0}(x)\mu^n_{\Omega_{0m}}(x)}
\nonumber\\
&=&
\frac{
<\Psi,[E(x)-E_n(x)]\Psi'>\mu_{\Omega_0}(x)+
\mu_{\Psi,\Psi'}(x) <\Omega,[E_n(x)-E(x)]\Omega_0>
}
{\mu_\Omega(x)[\mu_{\Omega_0}(x)+<\Omega,[E_n(x)-E(x)]\Omega_0>]}
\nonumber\\
&& +\frac{
[\mu^n_{\Psi-\Psi_m,\Psi'}(x)+\mu^n_{\Psi_m,\Psi'-\Psi'_m}(x)]
\mu^n_{\Omega_{0m}}(x)
+\mu^n_{\Psi_m,\Psi'_m}(x)[
\mu^n_{\Omega_{0m}-\Omega_0,\Omega_{0m}}(x)
+\mu^n_{\Omega_0,\Omega_{0m}-\Omega_0}(x)]
}
{[\mu_{\Omega_0}(x)+\mu^n_{\Omega_{0m}-\Omega_0,\Omega_{0m}}(x)
+\mu^n_{\Omega_0,\Omega_{0m}-\Omega_0}(x)+
<\Omega_0,[E_n(x)-E(x)]\Omega_0>]
}
\times \nonumber\\
&& \times \frac{1}{[\mu_{\Omega_0}(x)+<\Omega_0,[E_n(x)-E(x)]\Omega_0>]}
\ea
Since the finite linear combinations of Fock states are dense,
for given $\epsilon>0,\;\Psi,\Psi',\Omega_0$ we find some $m$ and
corresponding unit vectors $\Psi_m,\;\Psi_m,\;\Omega_{0m}$ such that
$||\Psi-\Psi_m||,\;||\Psi'-\Psi'_m||,\;
||\Omega_0-\Omega_{0m}||<\epsilon$. Since $E_n(x),E(x)$ are projections
with
uniform operator norm at most unity we find, using the Schwarz inequality
e.g. $|\mu_{\Psi-\Psi_m,\Psi'}(x)|\le ||\Psi-\Psi_m||<\epsilon$ or
$|\mu^n_{\Psi-\Psi_m,\Psi'}(x)|\le ||\Psi-\Psi_m||<\epsilon$.
Next, for given $x>0,\;\epsilon>0,\;\Psi,\Psi',\Omega_0$ there exists
$n_0=n(x,\epsilon,\Psi,\Psi',\Omega_0)$ such that
$||[E_n(x)-E(x)]\Psi||,\;
||[E_n(x)-E(x)]\Psi'||,\;
||[E_n(x)-E(x)]\Omega_0||<\epsilon$ for all $n>n_0$. It follows e.g.
$|<\Psi,[E_n(x)-E(x)]\Psi'>|<\epsilon$. The absolute value of
(\ref{6.32b}) can therefore be estimated by
\ba \label{6.32c}
|\mbox{(\ref{6.32b})}|
&\le & \epsilon \{
\frac{\mu_{\Omega_0}(x)+|\mu_{\Psi,\Psi'}(x)|}
{\mu_{\Omega_0}(x)[\mu_{\Omega_0}(x)-\epsilon]}
\nonumber\\
&& +2\frac{\mu^n_{\Omega_{0m}}(x)+|\mu^n_{\Psi_m,\Psi'_m}(x)|}
{[\mu_{\Omega_0}(x)-\epsilon]
[\mu_{\Omega_0}(x)-3\epsilon]} \}
\nonumber\\
&\le&
\frac{2\epsilon}{\mu_{\Omega_0}(x)-\epsilon}
(\frac{1}{\mu_{\Omega_0}(x)}
+\frac{1}{\mu_{\Omega_0}(x)-3\epsilon})
\nonumber\\
&\le&
\frac{4\epsilon}{[\mu_{\Omega_0}(x)-3\epsilon]^2}
\ea
where we have assumed that, given $x>0,\Omega_0$ we have
$3\epsilon<\mu_{\Omega_0}(x)$ and used
$|\mu_{\Omega_0}(x)|,\;|\mu_{\Psi,\Psi'}(x)|\le 1$.
It is clear that given any $\delta>0$ we may choose $\epsilon$ such that
the last line of (\ref{6.32c}) can be made smaller than $\delta$.

We now wish to to calculate the approximation
\be \label{6.32d}
\frac{\mu^n_{\Psi_m,\Psi'_m}(x)}{\mu^n_{\Omega_{0m}}(x)}
\ee
to (\ref{6.32a}) more explicitly. Let
$\Omega^n_{N\alpha}:=\Omega^n_N\otimes e^n_\alpha$ where
$(\Omega^n_N)_N$ is a cyclic system for $\MCO_n$ on the Hilbert space
${\cal F}^\parallel_n$ which is the completion of the finite linear span
of Fock states with excitations at most in the first $n$ degrees of
freedom and $e^n_\alpha$ is a Fock basis of the orthogonal complement
$({\cal F}^\parallel_n)^\perp$. We may identify $\Omega^n_N$ with
$\Omega^n_{N0}$ where $e_0=1$ is the vacuum of
$({\cal F}^\parallel_n)^\perp$. A given vector $\Psi$ can then be written
in the form
$\Psi=\sum_{N,\alpha} \Psi^n_{N\alpha}(\MCO_n)\Omega^n_{N\alpha}$
for certain measurable functions $\Psi^n_{N\alpha}$.

Up to this point $m,n$ are uncorrelated. Choose $n\ge m$. Then
the measurable functions
$\Psi^n_{mN\alpha},\; \Psi^{n\prime}_{mN\alpha}, \Omega^n_{0mN\alpha}$
corresponding to $\Psi_m,\;\Psi'_m,\;\Omega_{0m}$ respectively are in fact
polynomials and moreover they vanish unless $\alpha=0$. Introducing
a direct integral representation of ${\cal F}^\parallel$ subordinate
to $\MCO_n$ based on the $\Omega^n_{N\alpha}$, formula (\ref{6.32d})
becomes
\be \label{6.32e}
\frac{\int_0^x \; d\mu^n(x) [\sum_N \rho^n_N(x)
\overline{\Psi^n_{mN}(x)}\;\Psi^{n\prime}{mN}(x)]}
{\int_0^x \; d\mu^n(x) [\sum_N \rho^n_N(x) |\Omega^n_{0mN}(x)|^2]}
\ee
where $\rho^n_N(x)$ are the Radon -- Nikodym derivatives of the spectral
measures $\mu^n_N(B):=<\Omega^n_{N0},E_n(B)\Omega^n_{N0}>$
with respect to the total measure
\be \label{6.32f}
\mu^n(B)=
\frac{\sum_N 2^{-N} \sum_\alpha 2^{-\alpha}
<\Omega^n_{N \alpha},E_n(B)\Omega^n_{N\alpha}>}{\sum_{N,\alpha}
2^{-(N+\alpha)}}
=\frac{\sum_N 2^{-N} \mu^n_N(B)}{\sum_N 2^{-N}}
\ee
where we exploited that $E_n(B)$ does not act on the $e_\alpha$.

In order to compute (\ref{6.32e}) in the limit $x\to 0$ we now will
use a convenient choice of $\Omega^n_N$. These are simple modifications
of the example of $n$ mutually commuting constraints that we discussed in
the companion paper \cite{II}. The Hilbert space ${\cal F}$ is unitarily
equivalent to the Hilbert space $L_2(S,d\mu_G)$ where $S$ is
the set of real valued sequences $(p_I)_{I=1}^\infty$ and $\mu_G$ is a
Gaussian measure with white noise covariance, that is
$\mu_G(\exp(iF_f))=\exp(-\frac{\alpha}{2}\sum_I f_I^2)$ where
$F_f=\sum_I f_I p_I$. This follows immediately from
\be \label{6.32g}
\mu_G(e^{iF_f}):=<\Omega,e^{-\sum_I f_I[\hat{z}^{(3)}_I-
(\hat{z}^{(3)}_I)^\dagger]}\Omega>
\ee
and by using the Baker -- Campbell -- Hausdorff formula. Here, as before,
$\Omega$ denotes the Fock vacuum which in the $p-$representation
is given by $\Omega(p)=1$. We now choose the vectors $\Omega^n_N$
according to the first example in \cite{II}. Introduce the variables
$y_I:=p_I \sqrt{Q_I},\;I=1,..,n$ so that $\MCO_n=\frac{1}{2}
\sum_{I=1}^n y_I^2$ and let $h^n_{s,l},\;l\in L_s$ be the
harmonic polynomials of degree $s=0,1,..$ in the variables $y_I$ with
respect to the Laplacian in the $y_I,\;I=1,..,n$. Here $L_s$ is a
finite set of indices which depends on $s$. We define
\be \label{6.32h}
\Omega^n_{N=(s,l)}(\{p_I\}):=\Omega^n_{sl}(\{p_I\}):=c^n_{sl}
h_{sl}(\{x_I\})
\exp(-\frac{1}{4}\sum_{I=1}^n y_I^2)
\exp(\frac{1}{4\alpha}\sum_{I=1}^n y_I^2/Q_I)
\ee
where $c^n_{ls}$ is a normalization constant which is unimportant
for what follows. As we showed in \cite{II},
the $\MCO_n^k \Omega^n_{ls}$ are dense in ${\cal F}^\parallel_n$.
Using that formally
$d\mu_G(p)=\prod_{I=1}^\infty
e^{-p_I^2/(2\alpha)}\;dp_I/\sqrt{2\pi\alpha}$,
we now compute the
corresponding spectral measures
\ba \label{6.32i}
\mu^n_{sl}(x) &=&
\frac{\int_S\;d\mu_G(p) \;\theta(x-\MCO_n)\; |\Omega^n_{sl}(p)|^2}
{\int_S\;d\mu_G(p) \; |\Omega^n_{sl}(p)|^2}
\nonumber\\
&=&
\frac{\int_{\Rl^n}\;\frac{d^n p}{\sqrt{2\pi\alpha}^n}\;
e^{-\frac{1}{2\alpha}\sum_{I=1}^n p_I^2} \;\theta(x-\MCO_n)\;
|\Omega^n_{sl}(p)|^2}
{      \int_{\Rl^n}\;\frac{d^n p}{\sqrt{2\pi\alpha}^n}\;
e^{-\frac{1}{2\alpha}\sum_{I=1}^n p_I^2} \; |\Omega^n_{sl}(p)|^2}
\nonumber\\
&=&
\frac{\int_{\Rl^n}\;d^n y\;
e^{-\frac{1}{2}\sum_{I=1}^n y_I^2}
\;\theta(x-\frac{1}{2}\sum_{I=1}^n y_I^2)\;
|h^n_{sl}(y)|^2}
{\int_{\Rl^n}\;d^n y\;
e^{-\frac{1}{2}\sum_{I=1}^n y_I^2} \;|h^n_{sl}(y)|^2}
\nonumber\\
&=&
\frac{\int_{\Rl^n}\;d^n y\;
e^{-\frac{1}{2}\sum_{I=1}^n y_I^2}
\;\theta(x-\frac{1}{2}\sum_{I=1}^n y_I^2)\;
|h^n_{s0}(y)|^2}
{\int_{\Rl^n}\;d^n y\;
e^{-\frac{1}{2}\sum_{I=1}^n y_I^2} \;|h^n_{s0}(y)|^2}
\nonumber\\
&=&
\frac{1}{\Gamma(s+n/2)}\int_{\Rl_+}\;dt\; \theta(x-t)\;
t^{s+n/2-1}\; e^{-t}
\nonumber\\
&=& \mu^n_{s0}(x)
\ea
where in the second step we have integrated out all but the first
$n$ degrees of freedom, in the second we changed variables to $y_I$
thereby cancelling the Jacobean $\prod_{I=1}^n 1/\sqrt{Q_I}$ in numerator
and denominator, in the third we used that for given $s$ there exists
an $SO(n)$ rotation which transforms $h_{s0}$ into $h_{sl}$ and the
rotation invariance of $d^n y,\;\MCO_n$ and after this the calculation is
exactly the same as in \cite{II}. The total measure therefore becomes
as in \cite{II}
\be \label{6.32j}
\mu^n(x)=\frac{1}{2}\sum_{s=0}^\infty 2^{-s} \mu_{s0}(x)=
\frac{1}{2\Gamma(n/2-1)}\int_{\Rl_+}\;dt\; \theta(x-t)\;
t^{n/2-2}\; e^{-t/2}
\ee
and the Radon -- Nikodym derivatives were calculated to be
\be \label{6.32k}
\rho^n_{sl}(x)=\frac{d\mu^n_{sl}(x)}{d\mu^n(x)}=
\frac{2\Gamma(n/2-1)}{\Gamma(n/2+s)}\;
x^{s+n/2-1}\; e^{-x/2}\;[\int_0^x \; dt\; e^{-t/2}\; t^{n/2-2}]^{-1}
\ee
which close to $x=0$ behave as
\be \label{6.32l}
\frac{2\Gamma(n/2-1)}{\Gamma(n/2+s)}\;x^s
[(s+m/2-1)-x/2]\approx 2\delta_{s,0}
\ee
Notice that $L_0=\{0\}$ so there is only one harmonic polynomial of degree
zero. Inserting this into (\ref{6.32e}), formula (\ref{6.32d})
becomes
\be \label{6.32m}
\frac{\int_0^x \; d\mu^n(x) [\sum_{sl} \rho^n_{sl}(x)
\overline{\Psi^n_{msl}(x)}\;\Psi^{n\prime}_{msl}(x)]}
{\int_0^x \; d\mu^n(x) [\sum_{sl} \rho^n_{sl}(x) |\Omega^n_{0msl}(x)|^2]}
\ee
which close to zero becomes (the measures $\mu([0,x])$ cancel
both in numerator and denominator)
\be \label{6.32n}
 \frac{\sum_{sl} \rho^n_{sl}(0)
\overline{\Psi^n_{msl}(0)}\;\Psi^{n\prime}{msl}(0)}
{\sum_{sl} \rho^n_{sl}(0) |\Omega^n_{0msl}(0)|^2}
=\frac{\overline{\Psi^n_{m00}(0)}\;\Psi^{n\prime}{m00}(0)}
{|\Omega^n_{0m00}(0)|^2}
\ee
up to terms which vanish in the limit $x\to 0$ by the intermediate value
theorem of Lebesgue integral calculus (remember that the $\Psi^n_{msl}$
etc. are polynomials and non -- vanishing for finitely many $s,l$ only,
hence the integrand is actually a smooth function and $d\mu^n(x)=dx
\sigma^n(x)$ for a smooth function $\sigma_n$ is absolutely continuous
with respect to Lebesgue measure as follows from
(\ref{6.32k})).
Thus, (\ref{6.32n}) approximates (\ref{6.32a}) as closely as we want,
that is, for all
$x>0,\delta>0,\Psi,\Psi',\Omega_0$ we find
$m>m_0(x,\delta,\Psi,\Psi',\Omega_0)$ and
$n>n_0(x,\delta,\Psi,\Psi',\Omega_0)$ with $n\ge m$ such that these two
quantities differ at most by $\delta$.
Since (\ref{6.32n}) depends only on the coefficient $\Psi^n_{m00}(0)$
for any $m,n$ it follows that the physical Hilbert space in the
approximation given by $\delta$
in one to one correspondence with the one dimensional span of the vector
$\Omega^n_{m;s=0,l=0,\alpha=0}$, that is with $\Cl$. Since this holds for
all $\delta$, the physical Hilbert space selected by $\MCO$ from
${\cal F}^\parallel$ coincides with $\Cl$.
Since ${\cal F}={\cal F}^\parallel \otimes {\cal F}^\perp$ it follows
${\cal H}_{phys}={\cal F}^\perp$ as expected.\\
\\
Notice that this result is independent of the concrete choice of $Q_I$,
that is, independent of the choice of trace class operator.

\section{Linearized Gravity}

In this section we will consider the constraints of linearized gravity on Minkowski background. Because of the Minkowski background we can apply the same techniques as for the Maxwell-Gauss constraint.

 We will work with the  (real) connection formulation of canonical 
 gravity, i.e. the canonical pair of fields $(A^j_a,\tfrac{1}{\kappa 
\beta}E^a_j)$ on a three-dimensional manifold $\Sigma$, where $A^j_a$
is an $su(2)$-connection and $E^a_j=\sqrt{\text{det}(q)}e^a_j$ is a 
densitized triad for the spatial metric $q_{ab}$.
We use $a,b,c,\ldots$ for spatial indices and $i,j,k,\ldots$ for 
$su(2)$-indizes.
The latter ones are raised and lowered with $\delta^{ik}$ and 
$\delta_{ik}$, so we do not worry about the position of these indices. 
$\kappa=8\pi G_{Newton}$ denotes the gravitational coupling constant and $\beta$ the (real) Immirzi parameter. Canonical gravity in this formulation is a first class constraint system. The constraints of the full non-linear theory are the Gauss, the diffeomorphism and the scalar constraint:
\ba \label{lg1}
G_j &=&\partial_a E^a_j+\eps_{jkl}A^k_a E^a_l \nn \\
V_a &=& F^j_{ab}E^b_j  \nn \\
C &=& \frac{1}{\kappa \sqrt{\text{det}(q)}}\big(F^j_{ab}-(1+\beta^2)\eps_{jmn}K^m_a K^n_b \big)\eps_{jkl}E^a_k E^b_l  \q\q.
\ea
In the last line $K^j_a$ is given by $\beta \,K^j_a=A^j_a-\Gamma^j_a$, where $\Gamma^j_a$ is the spin connection of the triad $e^a_j$. $F=2(dA+A\wedge A)$ is the field strenght and the curvature of the connection $A$.

We assume that the fields are asymptotically flat and adopt the boundary conditions from \cite{Thasymp}, that is $A^j_a$ (and $K^j_a$) falls off as $r^{-2}$ and $E^a_j \sim {}^{\text{flat}}E^a_j+O(1/r)$ at infinity, where ${}^{\text{flat}}E^a_j$ is a densitized triad for a flat metric.

Linearized Gravity in the connection formulation was previously considered in \cite{AshRovLinGr}, however there complex connection variables ${}^\Cl A_a^j=\Gamma^j_a+iK^j_a$ were used. Since we have $\beta \,K^j_a=A^j_a-\Gamma^j_a$, we can express the complex connection in terms of the real variables as
\be \label{realcomplconn}
{}^\Cl A_a^j=\frac{i}{\beta}A_a^j+(1-\frac{i}{\beta})\Gamma_a^j \q.
\ee
Moreover \cite{AshRovLinGr} gave the ADM-energy functional in terms of the complex connection as
\be \label{ADMcompl}
H=\frac{1}{\kappa}\int_\Sigma d^3 x\, \sqrt{\text{det}(q)} \big(\overline{{}^\Cl A_a^{\; b}} {}^\Cl A_b^{\; a}- \overline{{}^\Cl A_a^{\; a}}{}^\Cl A_b^{\; b}\big)
\ee
where ${}^\Cl A_{ab}={}^\Cl A_a^{ j}e_b^j$. This energy functional can be derived as the surface term one has to add to make the integrated Hamiltonian constraint $C(N)=\int_\Sigma d^3 x N C$ (where $N$ is the lapse function) differentiable. The differentiability is achieved with respect to the fields $A_a^j,E^a_j$ fulfilling the asymptotic flat boundary conditions.

Rewriting the energy functional (\ref{ADMcompl}) with help of (\ref{realcomplconn}) into the real connection variables gives
\ba \label{ADMreal}
H=\frac{1}{\beta^2\kappa}\int_\Sigma d^3x \sqrt{\text{det}(q)}\big[ (1\!+\!\beta^2)[\Gamma_a^{\;b}\Gamma_b^{\;a}-\Gamma_a^{\;a}\Gamma_b^{\;b}]+[A_a^{\;b}A_b^{\;a}-A_a^{\;a}A_b^{\;b}]-2[\Gamma_a^{\;b}A_b^{\;a}-\Gamma_a^{\;a}A_b^{\;b}]\big]\;
\ea
where we introduced $A_{ab}={}^\Cl A_a^{ j}e_b^j$ and $\Gamma_{ab}=\Gamma_a^{ j}e_b^j$.

We come now to the linearization of the theory around Minkowski initial data $A^j_a=0$ and $E^a_j=\delta^a_j$. We will apply the following theorem, proved in a more general setting in \cite{AshLee}, which deals with the linearization of a finite dimensional first class constrained systems:
\begin{Theorem}
Let $(M,\Omega)$ be a finite dimensional symplectic manifold with coisotropic constraints $C_\alpha$ and Hamiltonian $H$. Suppose that $m_0 \in M$ is a point on the constraint surface at which the Hamiltonian vectorfield of $H$ vanishes. Consider the tangent space $V:=T_{m_0}(M)$ and introduce the linearized constraints, quadratic Hamiltonian, and linearized symplectic structure respectively by
\ba
&&  C^{\text{lin}}_\alpha:V \rightarrow C^\infty(V);\q\q
x \mapsto C^{\text{lin}}_\alpha
:=\frac{\partial C_\alpha}{\partial m^a}(m_0)x^a
\nn \\
&&   H^{\text{lin}}:V \rightarrow C^\infty(V);\q\q
x\mapsto H^{\text{lin}}
:=
H(m_0)+\frac{1}{2} \frac{\partial^2 H}{\partial m^a \partial m^b}(m_0)x^a x^b
\nn \\
&& \q\q\q\q\q\q\q\q \q\q\q\q\q\!
   \Omega_{\text{lin}}^{ab}
   :=\{x^a,x^b\}_{\text{lin}}
   :=\{m^a,m^b\}(m_0)=\Omega^{ab}(m_0) \q.
\ea
Here $x^a$ are understood as coordinates of a vector fields $x$ in the 
linearized 
phase space $V$, given by the relation $x^a=x\cdot m^a$ where $m^a$ are  
coordinates on the phase space $M$ (seen as $C^\infty$-functions on $M$). 
The statement is:

The $C^{\text{lin}}_\alpha$ are coisotropic for the symplectic vector space $(V
,\Omega^{\text{lin}})$, even Abelian, and $H^{\text{lin}}$ is invariant under their Hamiltonian flow. Moreover, the linearization of the reduced theory coincides with the reduction of the linearized theory.
\end{Theorem}

We will use the prescription of the theorem for our infinte dimensional theory and check afterwards whether the statement holds also for this case.
Thus we take $m_0=(A^j_a=0,E^a_j=\delta^a_j)$ and consider the tangent space at this point, coordinatized by ${}^\text{lin}A^j_a$ and ${}^\text{lin}H^a_j$, so that we can write $A^j_a={}^\text{lin}A^j_a$ and $E^a_j=\delta^a_j+{}^\text{lin}H^a_j$. We insert the latter into (\ref{lg1}) and keep only terms of linear order in $({}^\text{lin}A^j_a,{}^\text{lin}H^a_j )$.
The result is
\ba \label{lg2}
G_a^{\text{lin}}&=&\partial_b{}^\text{lin} H_{ba}+\eps_{abc}{}^\text{lin}A_{cb} \\
V_a^{\text{lin}}&=&\partial_a{}^\text{lin} A_{bb}-\partial_b{}^\text{lin} A_{ab} \\
C^{\text{lin}}&=&\eps_{abc}\partial_a{}^\text{lin} A_{bc}
\ea
where we introduced ${}^\text{lin} A_a^{\;\; b}:={}^\text{lin}A^j_a\delta^b_j$ and ${}^\text{lin} H^a_{\;\; b}={}^\text{lin} H^a_j\delta^j_b$ and all indices are pulled with respect to the flat background metric $\delta_{ab}$. (Hence we will not worry about index positions.) 
The induced symplectic structure is $\{{}^\text{lin} H_{ab}(x),{}^\text{lin}A^{cd}(y)\}_\text{lin}=\beta\kappa\delta_a^c\delta^b_d \, \delta(x,y)$.

To compute the linearized Hamiltonian we notize that (\ref{ADMreal}) is at least quadratic in $\Gamma_{ab}$ and $A_{ab}$. So we need $\Gamma_{ab}$ in terms of $({}^\text{lin}A^j_a,{}^\text{lin}H^a_j )$ to linear order, which is
\be \label{lingravgammaH}
{}^\text{lin}\Gamma_a^{\; b}=\frac{1}{2}\eps^{bcd}(\partial_a {}^\text{lin}H_{cd}+\partial_d(\delta_{ac} {}^\text{lin}H_{ee}-{}^\text{lin}H_{ac}-{}^\text{lin}H_{ca})) \q.
\ee
 This can be computed by using the condition $0=\partial_a E^a_j+\eps_{jkl}\Gamma^k_a E^a_l$ for the spin connection $\Gamma^k_a$. We can then replace $\Gamma,A,\sqrt{\text{det}(q)}$ by ${}^\text{lin}\Gamma,{}^\text{lin}A,1$ respectively in order to arrive at the quadratic Hamiltonian
\ba \label{ADMlin}
H^\text{lin}=
\frac{1}{\beta^2\kappa}\int_\Sigma d^3x \bigg( (1\!+\!\beta^2)[{}^\text{lin} \Gamma_a^{\;b}\;{}^\text{lin} \Gamma_b^{\;a}-{}^\text{lin} \Gamma_a^{\;a}\;{}^\text{lin} \Gamma_b^{\;b}]+[{}^\text{lin} A_a^{\;b}\;{}^\text{lin} A_b^{\;a}-{}^\text{lin} A_a^{\;a}\;{}^\text{lin} A_b^{\;b}] \nn \\
-2[{}^\text{lin} \Gamma_a^{\;b}\;{}^\text{lin} A_b^{\;a}- {}^\text{lin}\Gamma_a^{\;a}\;{}^\text{lin} A_b^{\;b}]\bigg)\q . \q
\ea
Now it is merely a computational effort to check, that the linearized constraints (\ref{lg2}) are indeed Abelian and that the Hamiltonian (\ref{ADMlin}) is invariant (on the constraint hypersurface) with respect to the Hamiltonian flow of the linearized constraints.

Summarizing, we  use for the linearized gravitational field canonical variables $({}^\text{lin} A_{ab},{}^\text{lin} H^{ab})$ (and drop in the following the superfix `lin'), subject to the Abelian constraints (\ref{lg2}). The boundary conditions for the linearized fields can be read off from the boundary conditions for the non-linear theory to be $A_{ab}\sim r^{-2}$ and $H^{ab}\sim r^{-1}$.

As in the Maxwell case we will use Fock space methods to quantize this system, so we have to introduce a pair of complex conjugated fields whose smeared forms will be finally promoted into annihilation and creation operators. Our choice is
\be
z_{ab}=\frac{1}{\sqrt{2}\beta l_p}\bigg[ W^{-1/2}\cdot( A_{ab}+ \eps_{acd}\partial_c H_{db})-i\beta W^{1/2}\cdot H_{ab} \bigg] \q \text{and}\q \bar{z}_{ab}\q,
\ee
where we introduced the abbreviation $W=\sqrt{-\Delta}$. The non-vanishing Poisson brackets are $i\hbar \{\bar{z}_{ab}(x),z_{cd}(y)\}=\delta(x,y)\delta_{ac}\delta_{ab}$.

With this choice the Hamiltonian can be written as a manifestly positive function, as will be shown later.

Now we plug
\ba
H_{ab} &=&\frac{i l_p}{\sqrt{2}}W^{-1/2}\cdot(z_{ab}-\bar z_{ab})=:l_p \,W^{-1/2}\cdot p_{ab} \nn \\
A_{ab}&=&\frac{\beta l_p}{\sqrt{2}}W^{1/2}\cdot (z_{ab}+\bar z_{ab})-\frac{i l_p}{\sqrt{2}}  W^{-1/2} \cdot \eps_{acd}\partial_c(z_{db} -\bar z_{db}) \nn \\ &=:&\beta l_p\, W^{1/2}\cdot x_{ab}-l_p\,  W^{-1/2} \cdot \eps_{acd}\partial_c p_{db}
\ea
into the constraints (\ref{lg2}) arriving at
\ba  \label{lg3}
\tilde G_a & :=& G_a  -W^{-2}\cdot\partial_a C  - W^{-2}\cdot\partial_b \eps_{abc} V_c \nn \\
&=& \beta l_p\, W^{1/2} \cdot\big(\eps_{acb}-W^{-2}\cdot\eps_{dbc}\partial_a\partial_d+W^{-2}\cdot \eps_{adb}\partial_d\partial_c  \big)\cdot x_{bc}+ l_p\,W^{-1/2}\cdot\partial_b \delta_{ac} p_{bc} \nn \\
V_a &=& \beta l_p\, W^{1/2} \cdot \big( \partial_a \delta_{bc}-\partial_c \delta_{ab}\big)\cdot x_{bc}+l_p\,W^{-1/2}\cdot\big(\eps_{bdc}\partial_a\partial_d-\eps_{bda}\partial_c\partial_d\bigg)\cdot p_{bc} \nn \\
C &=& \beta l_p \,W^{1/2} \cdot\eps_{abc}\partial_a x_{bc}+ l_p\, W^{-1/2} \cdot\big(-W^2 \delta_{bc}-\partial_b\partial_c\big)\cdot p_{bc} \q\q
\ea
where we introduced a more convenient equivalent set of constraints.

For the definition of the (classical) Master Constraint we use the Hilbert spaces $\fh=L_2(\Rl^3,d^3x)$ and $\fh^3$, with inner product on $\fh^3$ being given by $<f'_a,f_a>_{\fh^3}:=\int d^3x  \delta^{ab} \bar f'_a f_b$. Later on we will also need $\fh^9$ with $<f'_{ab},f_{ab}>_{\fh^9}:=\int d^3x  \delta^{ac}\delta^{bd}\bar{f'}_{ab} f_{cd}$. (Here we deviate from the usual mathematical notation and write the indices in the inner product in order to keep track of the ordering of the indices. For $\fh^3$ this is not necessary, therefore the indices are sometimes omitted.) Now we can define the Master Constraint as
\ba \label{lg4}
\MC=\tfrac{1}{2}\big( <\tilde G,K_1 \cdot \tilde G>_{\fh^3}+<V,K_2 \cdot V>_{\fh^3}+<C,K_3 \cdot C>_\fh\big)
\ea
where $K_1,K_2$ are positive definite operators on $\fh^3$ and $K_3$ is a positive definite operator on $\fh$. These operators have to be chosen in such a way that the Master constraint is well defined (on the classical phase space) and can be promoted into a densely defined operator on the quantum configuration space. In the following we will rewrite the Master constraint using the fields $x_{ab}$ and $p_{ab}$, smeared with some basis of $\fh^9$. To begin with the construction of this basis we introduce an orthonormal basis $b_I$ of $\fh$ consisting of real valued smooth functions of rapid decrease. From this we define a basis for the longitudinal modes in $\fh^3$ by $b_{Ia}^{(3)}:=W^{-1}\cdot \partial_a b_I$ and complete these to a full orthonormal basis of $\fh^3$ by choosing  smooth, transversal functions $b_{Ia}^{(1)},b_{Ia}^{(2)}$ of rapid decrease. Finally we equip $\fh^9$ with the following basis (the index $i$ assumes values $i=1,2)$:
\begin{xalignat}{3} \label{lingravBasis1}
& b_{Ibc}^{(1i)}:=\big(\eps_{acb}+W^{-2}\cdot \eps_{adb}\partial_d\partial_c  \big)\cdot b_{Ia}^{(i)} &
  &\mbox{left long. right transv.}  \nn \\
& b_{Ibc}^{(3)}:=-W^{-1}\cdot\partial_b b_{Ic}^{(3)}=-W^{-2}\cdot\partial_b\partial_c b_{I}  &  &\mbox{left and right long.} \nn \\
& b_{Ibc}^{(4i)}:=W^{-1}\cdot\partial_c b_{Ib}^{(i)}
  & &\mbox{left transv. right long.} \nn \\
& b_{Ibc}^{(6)} := -2^{-1/2}W^{-1}\!\!\!\!\cdot (\partial_a\delta_{bc}-\partial_c\delta_{ab})b_{Ia}^{(3)}=-2^{-1/2}W^{-2}\!\!\!\!\cdot (\Delta\delta_{bc}-\partial_c\partial_b)b_{I}    &  &\mbox{symm. transv., trace part} \nn \\
& b_{Ibc}^{(7)}:= 2^{-1/2} W^{-2}\cdot \eps_{bdc}\partial_a\partial_d b_{Ia}^{(3)} =-2^{-1/2}W^{-1}\cdot\eps_{bdc}\partial_d b_{I}
 & &\mbox{antisymm. transv.}
\end{xalignat}
 We complete this system to an orthonormal basis of $\fh^9$ by choosing a
 basis $b_{Ibc}^{(8)},b_{Ibc}^{(9)}$ of the symmetric transverse traceless
 modes\footnote{The completeness of this basis is verified in appendix
\ref{appgrav}.}. Furthermore we introduce another basis of the left 
longitudinal right transversal and left transversal right longitudinal modes:
\begin{xalignat}{2}
& b_{Ibc}^{(1'i)}:=-W^{-1}\cdot\partial_b \delta_{ac}b_{Ia}^{(i)}=W^{-1}\cdot \eps_{cda}\partial_d b_{Iba}^{(1i)}  && \mbox{left long. right transv.}   \nn \\
& b_{Ibc}^{(4'i)}:=-W^{-2}\cdot \eps_{bda} \partial_c \partial_d b_{Ia}^{(i)}= -W^{-1} \cdot \eps_{bda}\partial_d b_{Iac}^{(4i)} \q\q\q  & &\mbox{left transv. right long.}  \q.
\end{xalignat}
Since the operator $\delta: t_c \mapsto \eps_{cda}\partial_d t_a$ squares to $W^2=-\Delta$ on the subspace of the transversal modes in $\fh^3$, one can also write
\ba
b_{Ibc}^{(1i)}&=& W^{-1}\cdot \eps_{cda}\partial_d b_{Iba}^{(1'i)} \nn \\
b_{Ibc}^{(4i)}&=& -W^{-1} \cdot \eps_{bda}\partial_d b_{Iac}^{(4'i)} \q.
\ea
If one chooses the basis of the transversal modes in $\fh^3$ such that $b^{(2)}_{Ic}=W^{-1} \eps_{cda}\partial_d b^{(1)}_{Ia}$ one obtains
\ba \label{lgbaspr}
& b_{Ibc}^{(1'2)}=b_{Ibc}^{(11)}  \q\q & b_{Ibc}^{(1'1)}=b_{Ibc}^{(12)}  \nn \\
& b_{Ibc}^{(4'2)}=-b_{Ibc}^{(41)}  \q\q & b_{Ibc}^{(4'1)}=-b_{Ibc}^{(42)}  \q.
\ea

Now we come to the calculation of the first term in the Master Constraint (\ref{lg4})
\ba
<\tilde G,K_1 \cdot \tilde G>_{\fh^3}&=& \sum_{I,J;i,j=1,2,3} <\tilde G_a,b_{Ia}^{(i)}>_{\fh^3}<b_{Ia}^{(i)},K_1\cdot b_{Ja}^{(j)}>_{\fh^3}< b_{Ja}^{(j)},\tilde G_a>_{\fh^3} \q .
\ea
Using the explicit form (\ref{lg3}) of the modified Gauss constriant we can write
\ba
<\tilde G_a,b_{Ia}^{(i)}>_{\fh^3}&=&\beta l_p <W^{1/2} \cdot\big(\eps_{acb}-W^{-2}\cdot\eps_{dbc}\partial_a\partial_d+W^{-2}\cdot \eps_{adb}\partial_d\partial_c  \big)\cdot x_{bc},b_{Ia}^{(i)}>_{\fh^3} \nn \\
&& +l_p<W^{-1/2}\cdot\partial_b \delta_{ac} p_{bc},b_{Ia}^{(i)}>_{\fh^3} \nn \\
&\underset{i=1,2}{=}&\beta l_p<x_{bc},W^{1/2} \cdot\big(\eps_{acb}+W^{-2}\cdot \eps_{adb}\partial_d\partial_c  \big)\cdot b_{Ia}^{(i)}>_{\fh^9} \nn \\
&& - l_p<p_{bc},W^{-1/2}\cdot\partial_b \delta_{ac}b_{Ia}^{(i)}>_{\fh^9}  \\
&=&\!\! l_p \sum_{J;j=1,2}\bigg(\beta <x_{bc},\big(\eps_{acb}+W^{-2}\cdot \eps_{adb}\partial_d\partial_c  \big)\cdot b_{Ja}^{(j)}>_{\fh^9} \nn \\
&& - <p_{bc},W^{-1}\cdot\partial_b \delta_{ac}b_{Ja}^{(j)}>_{\fh^9} \bigg)<W^{1/2}b_{Ja}^{(j)},b_{Ia}^{(i)}>_{\fh^3}
\nn \\
%
&=&\!\! l_p \sum_{J;j=1,2}\big(\beta <x_{bc},b_{Jbc}^{(1j)}>_{\fh^9}+<p_{bc},b_{Jbc}^{(1'j)}>_{\fh^9} \big)<W^{1/2}b_{Ja}^{(j)},b_{Ia}^{(i)}>_{\fh^3} \q  \nn
\ea
where in the second line we used that $b^{(1)}_{Ia},b^{(2)}_{Ia}$ are 
transversal and used an integration by parts. In the second last line we expanded the transversal vectors $W^{1/2}\cdot b_{Ia}^{(i)},\, i=1,2$ in terms of the basis $b_{Ja}^{(j)},\, j=1,2$.

In a similar way
\ba
<\tilde G_a,b_{Ia}^{(3)}>_{\fh^3}&=&l_p<p_{bc},W^{1/2}\cdot b_{Ibc}^{(3)}>_{\fh^9} \nn \\
&=& l_p\sum_{J}<p_{bc},b_{Jbc}^{(3)}>_{\fh^9}<W^{1/2}\cdot b_{Ja}^{(3)},b_{Ia}^{(3)}>_{\fh^3} \q.
\ea
Thus, utilizing
\ba
\sum_{I,J;i,j=1,2,3}<W^{1/2}b_{I'a}^{(i')},b_{Ia}^{(i)}>_{\fh^3}<b_{Ia}^{(i)},K_1\cdot b_{Ja}^{(j)}>_{\fh^3}<b_{Ja}^{(j)},W^{1/2}b_{J'a}^{(j')}>_{\fh^3} \nn \\
 =<b_{I'a}^{(i')},W^{1/2} K_1 W^{1/2} \cdot b_{J'a}^{(j')}>_{\fh^3}
\ea
we get
\ba
<\tilde G_a,K_1 \cdot \tilde G_a>_{\fh^3}&=&
 l_p^2 \sum_{I,J;i,j=1,2}
\big(\beta <x_{bc},b_{Ibc}^{(1i)}>_{\fh^9}+<p_{bc},b_{Ibc}^{(2i)}>_{\fh^9} \big)<b_{Ia}^{(i)},W^{1/2} K_1 W^{1/2} \cdot b_{Ja}^{(j)}>_{\fh^3} \nn\\
&&\q\q\q\q \big(\beta <b_{Jbc}^{(1j)},x_{bc}>_{\fh^9}+<b_{Jbc}^{(2j)},p_{bc}>_{\fh^9}\big) \nn \\
&& +l_p^2\sum_{I,J}<p_{bc},b_{Ibc}^{(3)}>_{\fh^9}<b_{Ia}^{(3)},W^{1/2} K_1 W^{1/2} \cdot b_{Ja}^{(3)}>_{\fh^3}<b_{Jbc}^{(3)},p_{bc}>_{\fh^9}
\ea
where we assume, that $K_1$ (and later also $K_2$) commutes with the projector on the transversal modes. We will see that this is always possible to achieve.

In an analogous way as for $\tilde G_a$ we get for the $V_a$ contribution
\ba
<V_a,K_2 \cdot V_a>_{\fh^3}&=&
 l_p^2\sum_{I,J;i,j=1,2}
\big(\beta <x_{bc},b_{Ibc}^{(4i)}>_{\fh^9}+<p_{bc},b_{Ibc}^{(4'i)}>_{\fh^9} \big)<b_{Ia}^{(i)},W^{3/2} K_2 W^{3/2} \cdot b_{Ja}^{(j)}>_{\fh^3} \nn\\
&&\q\q\q\q \big(\beta<b_{Jbc}^{(4j)},x_{bc}>_{\fh^9}+<b_{Jbc}^{(4'j)},p_{bc}>_{\fh^9}\big) \nn \\
&& +2l_p^2\sum_{I,J} \big(\beta <x_{bc},b_{Ibc}^{(6)}>_{\fh^9} +  <p_{bc},b_{Ibc}^{(7)}>_{\fh^9}\big) <b_{Ia}^{(3)},W^{3/2} K_2 W^{3/2} \cdot b_{Ja}^{(3)}>_{\fh^3}\nn\\
&&\q\q\q\q\q \big(\beta<b_{Jbc}^{(6)},x_{bc}>_{\fh^9}+<b_{Jbc}^{(7)},p_{bc}>_{\fh^9}\big)
\ea

To obtain the $C$ contribution we need
\ba
<C,b_I>_\fh &=&l_p <\beta\,W^{1/2} \cdot\eps_{abc}\partial_a x_{bc}+W^{-1/2} \cdot\big(-W^2 \delta_{bc}-\partial_b\partial_c\big)\cdot p_{bc},b_I> \nn \\
   &=& -\beta l_p<x_{bc},W^{1/2}\cdot \eps_{abc}\partial_a b_I>_{\fh^9}+l_p<p_{bc}, W^{-1/2} \cdot (-W^2 \delta_{bc}-\partial_b\partial_c)\cdot b_I>_{\fh^9} \nn \\
  &=& -2^{1/2}l_p\big(\beta <x_{bc},W^{3/2} b_{Ibc}^{(7)}>_{\fh^9} +<p_{bc},W^{3/2} b_{Ibc}^{(6)}>_{\fh^9}\big) \nn \\
 &=& -2^{1/2}l_p\sum_{J}\big( \beta<x_{bc}, b_{Jbc}^{(7)}>_{\fh^9} +<p_{bc}, b_{Jbc}^{(6)}>_{\fh^9} \big)<W^{3/2} b_{J},b_I>
\ea
arriving at
\ba
<C,K_3 C>_\fh &=&
\sum_{I,J} <C,b_I>_\fh<b_I,K_2 b_J>_\fh<b_J,C>_\fh \nn \\
&=& 2 l_p^2 \sum_{I,J} \big(\beta<x_{bc}, b_{Ibc}^{(7)}>_{\fh^9} +<p_{bc}, b_{Ibc}^{(6)}>_{\fh^9} \big)<b_I,W^{3/2}K_3 W^{3/2} \cdot b_J>_\fh \nn \\
&&\q\q \, \big( \beta <b_{Ibc}^{(7)},x_{bc}>_{\fh^9}+<b_{Ibc}^{(6)},p_{bc}>_{\fh^9} \big)
\ea

The expression of the constraints in terms of the fields $z_{ab}$ and $\overline{z}_{ab}$ is quite complicated, therefore we will perform a canonical transformation which will simplify the constraints. We will describe the canonical transformation in terms of the coordinates
\be
x_I^{(\alpha)}:=<b_{Ibc}^{(\alpha)},x_{bc}>_{\fh^9} \q \text{and} \q p_I^{(\alpha)}:=<b_{Ibc}^{(\alpha)},p_{bc}>_{\fh^9} \q.
\ee
We define new coordinates $\tilde{x}_I^{(\alpha)}$ and $\tilde{p}_I^{(\alpha)}$ by
\begin{xalignat}{2} \label{lingravcanontrafo}
&\tilde x_{I}^{(1i)}=\frac{1}{\sqrt{2}}(\beta x_{I}^{(1i)}+p_I^{(1'i)}) &
&\tilde p_{I}^{(1i)}=\frac{1}{\sqrt{2}}(\frac{1}{\beta} p_I^{(1i)}-x_I^{(1'i)}) \nn \\
&\tilde x_{I}^{(3)}=p_I^{(3)} &
&\tilde p_{I}^{(3)}=-x_I^{(3)} \nn \\
&\tilde x_{I}^{(4i)}=\frac{1}{\sqrt{2}}(\beta  x_{I}^{(4i)}+p_I^{(4'i)}) &
&\tilde p_{I}^{(4i)}=\frac{1}{\sqrt{2}}(\frac{1}{\beta} p_I^{(4i)}-x_I^{(4'i)}) \nn \\
&\tilde x_{I}^{(6)}=\frac{1}{\sqrt{2}}(\beta  x_{I}^{(6)}+p_I^{(7)}) &
&\tilde p_{I}^{(6)}=\frac{1}{\sqrt{2}}(\frac{1}{\beta} p_I^{(6)}-x_I^{(7)}) \nn \\
&\tilde x_{I}^{(7)}=\frac{1}{\sqrt{2}}(\beta  x_{I}^{(7)}+p_I^{(6)}) &
&\tilde p_{I}^{(7)}=\frac{1}{\sqrt{2}}(\frac{1}{\beta} p_I^{(7)}-x_I^{(6)}) \nn \\
&\tilde x_{I}^{(8i)}=x_{I}^{(8i)} && \tilde p_{I}^{(8i)}= p_{I}^{(8i)}
 \q.
\end{xalignat}
Here, for instance $x_I^{(1'i)}$ stands for
\be \label{lgcantr}
x_I^{(1'i)}:=<b_{Ibc}^{(1'i)},x_{bc}>_{\fh^9}=\sum_{J;j=1,2}<b_{Jbc}^{(1j)},x_{bc}>_{\fh^9}<b_{Ibc}^{(1'i)},b_{Jbc}^{(1j)}>_{\fh^9} \q,
\ee
and the sum over $J,j$ reduces to just one term if one uses a basis with the property (\ref{lgbaspr}). In this case the canonical transformation restricts to finite dimensional subspaces, indexed by $I$.

This gives new complex fields $\tilde{z}^{(\alpha)}_I$, defined by
\ba
\tilde{z}^{(\alpha)}_I=\frac{1}{\sqrt{2}}(\tilde x_{I}^{(\alpha)}-i p_{I}^{(\alpha)}  ) \q\q
\bar{\tilde{z}}^{(\alpha)}_I=\frac{1}{\sqrt{2}}(\tilde x_{I}^{(\alpha)}+i p_{I}^{(\alpha)}  ) \q.
\ea

The Master Constraint is then
\ba
\MC =\frac{l_p^2}{4}\Big( \sum_{I,J;i,j=1,2} Q_{1IJ}^{\;(ij)}\, {\tilde{x}}_I^{(1i)} {\tilde{x}}_J^{(1j)}+2 \sum_{I,J}Q_{1IJ}^{\;(33)}\, {\tilde{x}}_I^{(3)} {\tilde{x}}_J^{(3)}
+\sum_{I,J;i,j=1,2} Q_{2IJ}^{\;(ij)} \, {\tilde{x}}_I^{(4i)} {\tilde{x}}_J^{(4j)}   \nn \\
 + 2\sum_{I,J}Q_{2IJ}^{\;(33)}\, {\tilde{x}}_I^{(6)} {\tilde{x}}_J^{(6)}
+2\sum_{I,J}Q_{3IJ}\, {\tilde{x}}_I^{(7)} {\tilde{x}}_J^{(7)} \Big)
\ea
where $Q_{1IJ}^{(ij)},Q_{2IJ}^{(ij)}$ and $Q_{3IJ}$ are the matrix elements of $Q_1=W^{1/2}K_1 W^{1/2},\;Q_2=W^{3/2}K_2 W^{3/2}$ and $Q_3=W^{3/2}K_2 W^{3/2}$ respectively.

We come now back to our assertion that the ADM-energy functional can be written as a manifestly positive function in terms of the fields $z_{ab}$ and $\bar z_{ab}$. We will calculate the energy functional in a specific gauge and then generalize to the whole phase space. To this end we note that the constraint hypersurface is given by the vanishing of all the $\tilde x_I^{(\alpha)},\, I \in \ci$ where $\alpha$ runs through all the values $1i,3,4i,6,7$, that is all modes except the symmetric transverse traceless ones. An natural gauge condition is to require the vanishing of $\tilde p_I^{(\alpha)},\, I \in \ci$ where the index $\alpha$ runs through the same values as above. Thus in this gauge the fields $A_{ab}$ and $H_{ab}$ are symmetric transversal traceless covariant tensors of second rank. (Hence this gauge is called STT gauge.) In the STT gauge formula (\ref{lingravgammaH}) simplifies to
\be
\Gamma'_{ab}=-\eps_{acd}\partial_cH'_{db}
\ee
where the prime indicates that the fields are in the STT gauge. Using that $-\eps_{acd}\partial_c$ squares to $-\Delta$ on transversal fields the energy functional (\ref{ADMlin}) can be written as
\ba
H^\text{lin}&=&
\frac{1}{\beta^2\kappa}\int_\Sigma d^3x \bigg( \beta^2 \Gamma'_{ab}\Gamma'_{ab}+(A'_{ab}-\Gamma'_{ab}) (A'_{ab}-\Gamma'_{ab})\bigg) \nn \\
&=& \frac{1}{\beta^2\kappa}\int_\Sigma d^3x \bigg( \beta^2 H'_{ab} (-\Delta)H'_{ab}+(A'_ab + \eps_{acd} \partial_c H'_{db})(A'_ab + \eps_{aef} \partial_e H'_{fb})\bigg) \nn \\
&=& 2 \int_\Sigma d^3x\, \overline{z'}_{ab}(\hbar \sqrt{-\Delta} \cdot z'_{ab})\q\q.
\ea
Since the energy functional is a gauge-invariant function (on the constraint hypersurface) we can rewrite it in terms of the fields which are not necessary in the STT gauge by introducing the projector $P^{(8)}$ onto the symmetric transversal traceless modes (see (\ref{lingravBasis2})):
\be
H^\text{lin}=2 \int_\Sigma d^3x\, \overline{z}_{ab}(\hbar \sqrt{-\Delta}\cdot P^{(8)} \cdot z)_{ab}+\mbox{terms vanishing on the constraint surface.}
\ee
 The canonical transformation (\ref{lingravcanontrafo}) leaves the STT modes unaffected, hence one can express the ADM-energy functional equally well in the $\tilde{z}_{ab},\overline{\tilde{z}}_{ab}$ fields by simply replacing the old with the new fields. We conclude that the energy functional is a manifestly positive function.

We will now quantize linearized gravity and examine the conditions on the operators $K_i,i=1,2,3$. As in the Maxwell case we quantize the theory by introducing the kinematical algebra generated from
\be
\hat{\tilde{z}}^{(\alpha)}_I=\widehat{<b^{(\alpha)}_{Ibc},\tilde{z}_{bc}>_{\fh^9}}
\ee
and the corresponding adjoint operators $(\hat{\tilde{z}}^{(\alpha)}_I)^\dagger$. This algebra is represented on the Fock space $\Ff$ generated from the cyclic vacuum vector $\Omega$. The commutation relations are the usual ones
\ba
\big[\hat{\tilde{z}}^{(\alpha)}_I,\hat{\tilde{z}}^{(\gamma)}_J\big]=0 \q\q\q\big[(\hat{\tilde{z}}^{(\alpha)}_I)^\dagger,(\hat{\tilde{z}}^{(\gamma)}_J)^\dagger\big]=0 \q\q\q  \big[\hat{\tilde{z}}^{(\alpha)}_I,(\hat{\tilde{z}}^{(\gamma)}_J)^\dagger\big]=\delta_{IJ}\delta^{\alpha\gamma} \q.
\ea

The Master Constraint Operator is given by
\ba
\MCO =\frac{l_p^2}{4}\Big( &\sum_{I,J;i,j=1,2} Q_{1IJ}^{\;(ij)}\,\hat{\tilde{x}}_I^{(1i)}\hat{\tilde{x}}_J^{(1j)}+2\sum_{I,J}Q_{1IJ}^{\;(33)}\,\hat{\tilde{x}}_I^{(3)}\hat{\tilde{x}}_J^{(3)}
+\sum_{I,J;i,j=1,2} Q_{2IJ}^{\;(ij)} \,\hat{\tilde{x}}_I^{(4i)}\hat{\tilde{x}}_J^{(4j)}   \nn \\
 &+ 2\sum_{I,J}Q_{2IJ}^{\;(33)}\,\hat{\tilde{x}}_I^{(6)}\hat{\tilde{x}}_J^{(6)}
+2\sum_{I,J}Q_{3IJ}\,\hat{\tilde{x}}_I^{(7)}\hat{\tilde{x}}_J^{(7)} \Big)
\ea
where $Q_{1IJ}^{(ij)},Q_{2IJ}^{(ij)}$ and $Q_{3IJ}$ are the matrix elements of $Q_1=W^{1/2}K_1 W^{1/2},\;Q_2=W^{3/2}K_2 W^{3/2}$ and $Q_3=W^{3/2}K_2 W^{3/2}$ respectively.

A calculation completely analogous to (\ref{6.14}) reveals that $\MCO$ is a densely defined operator on $\Ff$ if and only if $Q_1$, $Q_2$ and $Q_3$ are trace class operators on $\fh^3$ respectively $\fh$.

One possible choice for these operators is
\ba \label{lgchoice}
Q_j&=&  P^{\perp} e^{- h} P^{\perp}+ P^{\parallel} e^{- h} P^{\parallel}  \q\q \text{for} \q j=1,2 \nn \\
Q_3&=&e^{- h}
\ea
where $ h$ is the three dimensional harmonic oscillator operator and $P^\perp,P^\parallel$ are the projectors onto the transversal and longitudianl modes in $\fh^3$ respectively.$K_1=P^{\perp}W^{-1/2} e^{- h}W^{-1/2} P^{\perp}+ P^{\parallel}W^{-1/2} e^{- h}W^{-1/2} P^{\parallel}$ and analogously $K_2$ commute with the projector $P^\perp$.

The following calculation shows that the classical Master Constraint is well defined with this choice of the $Q_i$'s. We consider only the part $<\tilde G_a,K_1 \cdot \tilde G_a>$ since the other terms can be treated in a similar way. As in the Maxwell case we will expand this term into coherent states $\psi_z$ (\ref{6.16}) for the three dimensional harmonic oscilator $h$. In the following calculation we will see the components $(P^\perp \cdot \tilde G)_a$ and $(P^\parallel \cdot \tilde G)_a$ for $a=1,2,3$ as scalar functions. To emphasise this, we will place a dot above the $a$:$\;\dot{a}$. We will also replace the label $z \in \Cl^3$ in $\psi_z$ by $z_{\dot{a}} \in \Cl^3, \dot{a}=1,2,3$. We can then write:
\ba \label{lingravabschatz}
<\tilde G_a, K_1 \cdot \tilde G_a>_{\fh^3}\!\!\!
 &=& \sum_{\dot{a}=1,2,3}\sum_{\gamma=\perp,\parallel} <  P^{(\gamma)} W^{-1/2} \tilde G_{\dot{a}}, e^{-h} P^{(\gamma)}W^{-1/2} \tilde G_{\dot{a}}>_\fh \nn \\
&=&\sum_{\dot{a}=1,2,3} \sum_{\gamma=\perp,\parallel }
 \int_{\Cl^3} \frac{d^3 z_{\dot{a}} d^3 \bar{z}_{\dot{a}}}{\pi^3}
 \int_{\Cl^3} \frac{d^3 z_{\dot{a}}' d^3 \bar{z}_{\dot{a}}'}{\pi^3}
<P^{(\gamma)} W^{-1/2} \tilde G_{\dot{a}}, \psi_{z_{\dot{a}}}>_\fh  \nn \\
 && \q\q\q\q \times <\psi_{z_{\dot{a}}},e^{-h}\psi_{z_{\dot{a}}'}>_{\fh}
 <\psi_{z_{\dot{a}}'},P^{(\gamma)} W^{-1/2} \tilde G_{\dot{a}}>_\fh \nn \\
 &=&\sum_{\dot{a}=1,2,3} \sum_{\gamma=\perp,\parallel }
 \int_{\Cl^3} \frac{d^3 z_{\dot{a}} d^3 \bar{z}_{\dot{a}}}{\pi^3}
 \int_{\Cl^3} \frac{d^3 z_{\dot{a}}' d^3 \bar{z}_{\dot{a}}'}{\pi^3}
< W^{-2} \partial_b \partial_b P^{(\gamma)} W^{-1/2} \tilde G_{\dot{a}}, \psi_{z_{\dot{a}}}>_\fh   \nn \\
&&\q\q\q\q \times <\psi_{z_{\dot{a}}},e^{-h}\psi_{z_{\dot{a}}'}>_\fh
 <\psi_{z_{\dot{a}}'},W^{-2} \partial_c \partial_c P^{(\gamma)} W^{-1/2} \tilde G_{\dot{a}}>_\fh \nn \\
&=&\!\!\!\!\sum_{\dot{a}=1,2,3} \sum_{\gamma=\perp,\parallel }
\!\!\int_{\Cl^3}\!\!\! \frac{d^3 z_{\dot{a}} d^3 \bar{z}_{\dot{a}}}{\pi^3}
\!\!\! \int_{\Cl^3} \!\!\!\frac{d^3 z_{\dot{a}}' d^3 \bar{z}_{\dot{a}}'}{\pi^3}
< W^{-\tfrac{1}{2}} \partial_b P^{(\gamma)} W^{-\tfrac{1}{2}} \tilde G_{\dot{a}}, W^{-\tfrac{3}{2}}\partial_b \psi_{z_{\dot{a}}}>_\fh \nn \\
&& \q\q\q\q \times  <\psi_{z_{\dot{a}}},e^{-h}\psi_{z_{\dot{a}}'}>_\fh
<W^{-\tfrac{3}{2}} \partial_c \psi_{z_{\dot{a}}'},W^{-\tfrac{1}{2}} \partial_c  P^{(\gamma)} W^{-\tfrac{1}{2}} \tilde G_{\dot{a}}>_\fh \nn \\
&\leq & \sum_{\dot{a}=1,2,3} \sum_{\gamma=\perp,\parallel }
   || W^{-1/2} \partial_b P^{(\gamma)}\, W^{-1/2} \tilde G_{\dot{a}} ||^2_{\fh^3}
   \int_{\Cl^3}\!\! \frac{d^3 z_{\dot{a}} d^3 \bar{z}_{\dot{a}}}{\pi^3}
    \int_{\Cl^3} \!\!\frac{d^3 z_{\dot{a}}' d^3 \bar{z}_{\dot{a}}'}{\pi^3}  \nn \\
&& \q\q\q\q \times  ||W^{-3/2} \partial_b \psi_{z_{\dot{a}}}||_{\fh^3}
 ||W^{-3/2} \partial_b \psi_{z_{\dot{a}}'}||_{\fh^3}
 |<\psi_{z_{\dot{a}}},e^{-h}\psi_{z_{\dot{a}}'}>_\fh|
\ea
The first factor in the last line can be simplified to
\be
|| W^{-1/2} \partial_b P^{(\gamma)}\, W^{-1/2} \tilde G_{\dot{a}} ||^2_{\fh^3}=||(P^{(\gamma)} \tilde G)_{\dot{a}} ||^2_{\fh}\leq || \tilde G_{\dot{a}} ||^2_{\fh}
\ee
since partial derivatives commute with the projectors $P^{(\gamma)}$.
The remaining integrals in (\ref{lingravabschatz}) are the same as in the sixth line  of (\ref{6.19}), so we can use the estimates (\ref{6.19}--\ref{6.23}), showing that the Master constraint is a well defined phase space function.

If we assume,as is the case for the $Q_i$'s in (\ref{lgchoice}), that it is possible to choose $b_I$ and $b_{Ia}^{(1)}, b_{Ia}^{(2)}$ such that $b_I$, $b_{Ia}^{(1)}, b_{Ia}^{(2)}$ and $b_{Ia}^{(3)}=W^{-1}\cdot\partial_a b_I$ are eigenstates of $Q_1,Q_2$ and $Q_3$,  we can write the Master Constraint Operator more compactly as
\ba
\MCO=\frac{l^2_p}{2} \sum_{I,\alpha} \, q_I^{(\alpha)} \, (\hat{\tilde{x}}_I^{(\alpha)})^2 \q\q,
\ea
 where the index $\alpha$ assumes values $\alpha=1i,3,4i,6,7$ with $i=1,2$
 and the $q_I^{\alpha}$ are positive eigenvalues (multiplied by two for
 $\alpha=3,6,7$) of the $Q_j$'s. Thus this Master Constraint Operator has
 the same structure as the Master Constraint Operator for the Maxwell
theory (\ref{6.14a}). Going through the same procedure as in this case,
see
section \ref{s6.2}, we see that the physical Hilbert space is unitarily equivalent to the Fock space generated from the vacuum $\tilde \Omega$  by applying just the creation operators generating symmetric--transverse--traceless modes.

\section{Conclusions}
\label{s9}

The basic lesson that we have learnt from the examples studied in the
present paper is that, surprisingly, the Master Constraint Programme
can cope with the UV singularities that one expects from squaring
operator valued distributions. Take the Maxwell case for simplicity.
The infinite number of classical Gauss constraints $G(x)=\partial_a
E^a(x)$ can classically be encoded in the single Master contraint
$\MC'=\frac{1}{2} \int d^3x G(x)^2$. This is classically well defined
because
$G(x)^2$ decays at infinity as $r^{-6}$ and, moreover, it has well defined
distributional second derivatives with respect to the phase space
differentiable structure.
However, quantum mechanically, $\MCO'=\infty$
is hopelessly divergent. This cannot be even repaired by subtracting an
infinite constant because $\MCO'$, in contrast to the electromagnetic
field energy $H=\frac{1}{2} \int d^3x \delta_{ab} \delta_{ab}
(E^a(x) E^b(x)+B^a(x) B^b(x))$ has a different structure in terms
of creation and annihilation operators. This is best seen by using the
Fourier transform of the annihilation operators
$\hat{z}_a(k)=\int d^3x/\sqrt{2\pi}^3 e^{-ik\cdot x} \hat{z}_a(x)$ where
$z_a(x)=(\root 4\of{-\Delta} A_a(x)-i(\root 4\of{-\Delta})^{-1}
E^a(x))/\sqrt{2\alpha}$. Then we get
\ba \label{9.1}
\hat{H}&=&\hbar\int\; d^3k\; ||k||\; (\delta^{ab}-\frac{k^a k^b}{||k||^2})
\;
[\hat{z}_a(k)\hat{z}_b(k)^\dagger
+\hat{z}_a(k)^\dagger \hat{z}_b(k)]
\\
\MCO'&=&\frac{\alpha}{2}\int\; d^3k\; ||k||\; k^a k^b\;
[\hat{z}_a(k)\hat{z}_b(k)^\dagger
+\hat{z}_a(k)^\dagger \hat{z}_b(k)
-\hat{z}_a(k) \hat{z}_b(-k)
-\hat{z}_a(k)^\dagger \hat{z}_b(-k)^\dagger]
\nonumber
\ea
Thus, the normal ordered expression $:\hat{H}:$ is densely defined and
positive while $:\MCO':$ is neither densely defined nor
(formally) positive.

We used the flexibility in the definition of the Master Constraint in
order to circumvent this problem. What we did is to introduce a
positive integral kernel $K(x,x')$ (that is, a positive operator on
the one particle Hilbert space $\mathfrak{h}=L_2(\Rl^3,d^3x)$)
and to define $\MCO=\int d^3x \int d^3x' K(x,x') G(x) G(x')$
which still classically encodes all the constraints. The idea is that
the kernel serves to smoothen the operator valued distributions
into well behaved operators. We found that the
necessary and sufficient condition for $\MCO$ to be densely defined is
that $K$ be trace class. For any such choice of $K$ the physical Hilbert
space selected by the Master Constraint from the full Fock space of all
modes was the correct one, the Fock space for the transversal modes.
Thus, the physical Hilbert space constructed is independent of the
concrete choice of $K$ which matches nicely with the fact that the
classical constraint surface defined by $\MC=0$ is independent of the
choice of $K$.\\
\\
\\
\\
{\large Acknowledgements}\\
\\
We would like to thank Jos\'e Velhinho for fruitful discussions.
BD thanks the German National Merit Foundation for financial support.
This research project was supported in part by
a grant from NSERC of Canada to the Perimeter Institute for Theoretical
Physics.

\begin{appendix}

\section{Completeness of the basis} \label{appgrav}

Here we will verify that (\ref{lingravBasis1}) is indeed a basis of the Hilbert space of square integrable covariant tensors of second rank $\fh^9=(L_2(\Rl^3))^9$. We will begin by establishing, that the modes listed there are complete. To this end we introduce the projector $p$ onto the transversal modes in $\fh^3=(L_2(\Rl^3))^3$
\be
(p\cdot v)_a:=p_a^b\cdot v_b:=(\delta_a^b+W^{-2}\cdot \partial_a \partial_b)\cdot v_b
\ee
and define the following projectors on $\fh^9$
\begin{xalignat}{2}\label{lingravBasis2}
&(P^{(1)}\cdot T)_{ab}=(\delta_a^c-p_a^c)\cdot p_b^d \cdot T_{cd} && \mbox{2 left long. right transv. modes}\nn \\
&(P^{(3)}\cdot T)_{ab}=(\delta_a^c-p_a^c)\cdot(\delta_b^d-p_b^d) \cdot T_{cd} && \mbox{1 left and right long. mode}\nn \\
&(P^{(4)}\cdot T)_{ab}=p_a^c\cdot (\delta_b^d-p_b^d) \cdot T_{cd} &&\mbox{2 left transv. right long. modes}\nn \\
&(P^{(6)}\cdot T)_{ab}= \tfrac{1}{2}p_{ab}\cdot p^{cd}\cdot T_{cd} && \mbox{1 symm. transv. trace part mode} \nn \\
&(P^{(7)}\cdot T)_{ab}=\tfrac{1}{2}(p_a^c\cdot p_b^d-p_b^c\cdot p_a^d)\cdot T_{cd} &&\mbox{1 antisymm. transv. mode} \nn \\
&(P^{(8)}\cdot T)_{ab}= \tfrac{1}{2}(p_a^c\cdot p_b^d +p_b^c\cdot p_a^d-p_{ab}p^{cd})\cdot T_{cd} &&\mbox{2 symm. transv. tracefree modes}
\end{xalignat}
 Using the projector property $p\cdot p=p$, it is easy to see that the 
 projectors $P^{(\alpha)}$ are orthogonal to each other and satisfy
 $P^{(\alpha)}\cdot P^{(\beta)}=\delta^{\alpha\beta} P^{(\alpha)}$. For 
 instance
\ba
(P^{(6)}\cdot P^{(8)}\cdot T)_{ab}&=&\tfrac{1}{2}p_{ab}\cdot p^{ef}\cdot\tfrac{1}{2}(p_e^c\cdot p_f^d +p_f^c\cdot p_e^d-p_{ef}p^{cd})\cdot T_{cd} \nn \\
&=&\tfrac{1}{4}p_{ab}(p^{cd}+p^{cd}-p^{ef}\delta_{ef}p^{cd})=0
\ea
where we used $p^{ef}\delta_{ef}=2$. Furthermore the sum of all the projectors in (\ref{lingravBasis2}) is the identity on $\fh^9$.

Next we have to show that the $b_{I}^{(\alpha)}$ in (\ref{lingravBasis1}) 
are a complete basis of the subspace of $\alpha$-modes in $\fh^9$. We 
begin with the left longitudinal and right transversal modes, rewriting them as follows
\ba
b_{Ibc}^{(1i)}&:=&\big(\eps_{acb}+W^{-2}\cdot \eps_{adb}\partial_d\partial_c  \big)\cdot b_{Ia}^{(i)}
=-W^{-2} \cdot \partial_b \eps_{cda}\partial_d b_{Ia}^{(i)} \q.
\ea
This equation can be proved by using that every transversal vector $t_{1a}$ can be written as the curl operator applied to another transversal vector $t_{1a}=\eps_{abc}\partial_b t_{2c}$. Additionally one utilizes that $\eps_{abc}\partial_b \cdot \eps_{cde}\partial_d=-\Delta \,\delta_{ae}$ on the subspace of transversal vectors. Now we have for any left longitudinal tensor $T_{ab}$
\ba
T_{ab}=(\delta_a^c-p_a^c)\cdot T_{cb}=W^{-1} \partial_a \cdot(-W^{-1}\cdot\partial_c T_{cb})=:W^{-1}\cdot \partial_a f_b
\ea
so it can be written as a gradient of a vector $W^{-1}\cdot f_b$. If 
$T_{ab}$ is additionally right transversal this vector has to be 
transversal. Therefore $\{ b_{Jcd}=W^{-1} \cdot \partial_at_{dJ}\}_{j\in 
\Fj}$ is a(n) (orthonormal) basis for the left longitudinal right transversal modes if $\{t_{dJ}\}_{J\in \ci}$ is a(n) (orthonormal) basis for the transversal modes in $\fh^3$. Now if the latter is the case then $\{t'_{aJ}:=W^{-1}\cdot \eps_{abd} \partial_b t_{dJ}\}_{J\in \Fj}$ is also an orthonormal basis for the transversal vector modes, since the operator $t_a \mapsto W^{-1}\cdot \eps_{abd} \partial_b t_{a}$ is unitary on this subspace (on the transversal modes it squares to the identity and on the whole space $\fh^3$ to the projector $p$). So we have proved, that $b^{(1i)}$ and $b^{(1'i)}$ is a basis for the left longitudinal right transversal modes. In a similar way one can deal with $\alpha=3,4i,4'i$.

For the completeness proof of
\be
b^{(6)}_{Ibc}= -2^{-1/2}W^{-1}\cdot (\partial_a\delta_{bc}-\partial_c\delta_{ab})b_{Ia}^{(3)}\underset{b_{Ia}^{(3)}=W^{-1}\cdot \partial_a b_I}{=}2^{-1/2}(\delta_{bc}+W^{-2}\cdot\partial_c\partial_b)b_I^{(3)}
\ee
we write for a tensor $T_{ab} \in P^{(6)}(\fh^9)$:
\ba
 T_{ab}=(P^{(6)}\cdot
 T)_{ab}=\tfrac{1}{2}p_{ab}p^{cd}T_{cd}=\tfrac{1}{2}(\delta_{ab}+W^{-2}\cdot
 \partial_a \partial_b)\cdot T_{cc}
\ea
 where in the second equation we used that $T_{ab}$ is right and left
 transversal. $T_{cc}$ is a square integrable function, so it can be
 expanded into the basis $\{b_I\}_{I\in \Fj}$ of $\fh$. We conclude, that
 $\{b_{Ibc}^{(6)}\}_{I\in \Fj}$ is a complete basis of the symmetric and
transversal modes with tracepart.

 To verify the completeness of the antisymmetric and transversal basis
 $b_{Ibc}^{(7)}=-2^{-1/2}W^{-1}\cdot\eps_{bdc}\partial_d b_I$, we use that
every antisymmetric tensor $T_{ab}$ can be expressed as
\be
T_{ab}=\eps_{abc}t_c \q \mbox{where} \q t_c=\tfrac{1}{2}\eps_{cde} T_{de} \q.
\ee
 The transversality condition $\partial_a T_{ab}=0$ translates into the
 vanishing of the curl $\eps_{abc}\partial_a t_c$ of $t_c$, so that $t_c$
 has to be longitudinal, i.e. can be written as the gradient of a function
 $W^{-1} \cdot f$. If $T_{ab}$ is square integrable, so is $t_c$ and hence
 $f$, so $f$ can be expanded into a basis $b_I$ of $\fh$. We conclude,
 that $\{b_{Ibc}^{(7)}\}_{I\in \Fj}$ is a complete basis of the
antisymmetric and transversal modes.

This finishes our verification, that the basis (\ref{lingravBasis1}) is
complete.

\end{appendix}

\end{document}